\documentclass[11pt,a4paper]{article}
\usepackage{amsmath}
\usepackage{enumerate}
\usepackage{amssymb}
\usepackage{graphicx}
\usepackage{multirow}
\usepackage{xcolor}
\usepackage{tablefootnote}
\usepackage{threeparttable}
\usepackage{bm}
\usepackage[normalem]{ulem}
\usepackage{placeins}
\usepackage{jcappub}

\begin{document}

\title{The Age of the Universe with Globular Clusters IV: Multiple Stellar Populations}

\author[a]{David Valcin,}
\author[b,c]{Raul Jimenez,}
\author[d,e]{Carmela Lardo,}
\author[a,f]{Uro\v{s} Seljak,}
\author[b,c]{Licia Verde}

\affiliation[a]{Berkeley Center for Cosmological Physics and Department of Physics, University of California, Berkeley, CA 94720, USA}
\affiliation[b]{ICC, University of Barcelona, Mart\' i i Franqu\` es, 1, E08028
Barcelona, Spain}
\affiliation[c]{ICREA, Pg. Lluis Companys 23, Barcelona, 08010, Spain.} 
\affiliation[d]{Dipartimento di Fisica e Astronomia ``Augusto Righi Universit\`a di Bologna, via Piero Gobetti 93/2, I-40129 Bologna, Italy.}
\affiliation[e]{INAF - Osservatorio di Astrofisica e Scienza dello Spazio di Bologna, via Piero Gobetti 93/3, I-40129 Bologna, Italy.}
\affiliation[f]{Lawrence Berkeley National Lab, 1 Cyclotron Road, Berkeley, CA 94720, USA}

\emailAdd{dvalcin@berkeley.edu}
\emailAdd{raul.jimenez@icc.ub.edu}
\emailAdd{carmela.lardo2@unibo.it}
\emailAdd{useljak@berkeley.edu}
\emailAdd{liciaverde@icc.ub.edu}

\abstract{
We revisit the determination of the age of the Universe from galactic globular clusters, extending previous analyses by explicitly accounting for the presence of multiple stellar populations within each cluster. Using high--quality \textit{Hubble Space Telescope} color--magnitude diagrams for 69 globular clusters, we relax the standard single--population assumption, and model two stellar populations with independent ages, metallicities, helium abundances, and population fractions. The inference is performed using the full color--magnitude diagram morphology, an explicit treatment of field contamination, and a hierarchical framework that propagates non--Gaussian age posteriors. Allowing for multiple stellar populations has a negligible impact on globular cluster age estimates. The ages of the oldest populations remain fully consistent with those obtained under the single--population assumption, with differences at the $0.6\sigma$ level. Restricting to the metal--poor subsample ([Fe/H] $< -1.5$), we infer a dominant old component with mean age $t_{\rm GC}=13.61\pm0.25\,\mathrm{(stat)}\,\pm0.23 \mathrm{(sys)}\,\mathrm{Gyr}$. Adopting a conservative delay between the Big Bang and the formation of the first globular clusters, we obtain an age of the Universe of $t_{\rm U}=13.81\pm0.25\,\mathrm{(stat)}\,\pm0.23 \mathrm{(sys)}\,\mathrm{Gyr}$. In addition to age constraints, our analysis yields simultaneous measurements of metallicity and helium content for the different populations, including constraints on helium enrichment and population fractions which are consistent with independent determinations from the literature. These results demonstrate that globular--cluster--based cosmic chronometry is robust to stellar population complexity, reinforcing its role as a precise and largely cosmological model--independent probe of the age of the Universe.
}

\maketitle

\section{Introduction}
\label{sec:intro}

Cosmology seems to be at the crossroad: the extremely successful Lambda Cold Dark Matter (LCDM) model, under the pressure of two decades of precision cosmology and massive observational efforts of large surveys (Cosmic Microwave Background, galaxy redshift surveys, weak lensing surveys, supernovae surveys)  seems, possibly, to be showing some cracks. The Hubble tension has reached 6 sigma significance level (see, e.g., \cite{VerdeARAA}) and the latest DESI results (\cite{DESI2025} DESI2025) seem to favor an evolving dark energy. A recently-added pressure point consists in constraining the look back time using measurements of the ages of old cosmic objects. 

Galactic globular clusters (GC) are prime candidates to find the oldest objects in the Universe. It is well known that information about their age can be extracted from the full color magnitude diagram (CMD) provided other properties (metallicity [Fe/H], alpha enhancement [$\alpha$/Fe], distance, dust attenuation, reddening and primordial helium abundance) are determined simultaneously, and then effectively treated as nuisance parameters.  This is achieved by including only stars in hydrostatic equilibrium in the main sequence, sub-giant and red giant branches.  These are the easiest objects to model as they consist of a radiative core and a fully convective envelope, for which stellar evolution models have reached maturity. 

Because the GC age determination is independent of assumptions on the cosmological model, it can be used to constrain it. 
GC ages are independent of the early-Universe physics (on which probes like cosmic microwave background (CMB) and baryon acoustic oscillations  heavily rely on), and independent of very late-time physics (such as a possibly evolving  dark energy  or the physics assumptions that go into the classic cosmic distance ladder). 
GCs are actually a middle-redshift probe. GCs have been observed directly by JWST at $z > 10$ (Adamo et al. 2024) indicating that they must have been in place before then. The redshift range $1 < z < 20 (10)$ contains 50\% (40\%) of the age of the Universe;  $z < 0.2$ ($z<0.05$)  only carries information of 20\% (5\%) of the age of the Universe. In this sense, one could argue the oldest  GC ages carry (integrated) information on the Universe history from their formation time  (z$\sim$ 10 or 20) to their observation. This makes them highly complementary for example to cosmological Type 1a supernovae  that carry integrated information from $z=0$ to their redshift of observation. GC ages do not depend on cosmological distances nor on the distance ladder:  they only rely on stellar nuclear reactions as an atomic clock providing an absolute age determination.

A recent paper (\cite{Valcin25}, hereafter V25; see also previous works~\cite{Valcin2020,Valcin2021}),  using  the full CMD to constrain GCs parameters, presented age determinations of 69 GCs observed in HST optical filters F606W and F814W, obtaining a 5\% error on individual ages for the oldest GCs.  Taken together, these objects imply a Universe 13.57 ($\pm 3$\%) Gyr old. This represents a pressure point for $\Lambda$CDM. In fact, the age of the Universe in $\Lambda$CDM is  $\sim  13.857$ Gyr $\times  (68$/H$_0$[km/s/Mpc]) thus  the age of the Universe estimated from GC is in agreement with the early Universe (CMB)  predictions in $\Lambda$CDM. But for an H0 of $73$ km/s/Mpc, preferred by the Cepheids-based cosmic distance ladder, the Universe is expected to be almost a full Gyr (i.e. 7\%) younger than predicted by the CMB and $2\sigma$ younger than the age favored by the GC analysis. Of course, two sigma “does not a tension make”.

How sharp is this  ``pressure point"  really? 
Could it appear artificially sharp due to some of the modeling assumptions?  Ref.~\cite{Valcin2020} marginalized over distance, reddening, metallicity imposing priors motivated by independent observations and estimated possible systematics from stellar modeling. Subsequently, \cite{Valcin2021}, improved the metallicity determination, breaking the degeneracy between  metallicity and depth of the convection  envelope  thus reducing the systematic error in the age determination. 

In this paper, we relax the assumption that the GC is made of a single population. It is known that GCs have multiple stellar populations (see e.g.~\cite{ReviewBastian18}). We increase the parameter space by allowing two population with independent values for the age and metallicity.
In section 2 we present the modifications to the methodology of V25 introduced to model multiple populations in section 3 we present the  resulting constraints on the age of the clusters and the Universe while in section 4 we report the  constraints in  the other clusters properties  and compared them with  existing literature. Finally we conclude in secion 5.

\section{Data and Methodology}

The data and methods used  follow closely 
V25, where  detailed descriptions  can be found.
The data  set is the same: 69 galactic GC observed in Vega filters F606W and F814W \cite{Valcin2020, Valcin2021} and refs. therein. 
The methodological modifications for the present study only touch two aspects: improved photometric cleaning, including empirical modeling of field contamination, and allowing for the presence of two independent populations each with  their own age, metallicity and helium content. This affects the ridgeline extraction method, the form of the likelihood and thus the details of the hierarchical likelihood analysis. 
We describe the modifications in detail below and refer the reader to V25 for the  description of the original methodology.

\subsection{Updated photometric cleaning}
\label{sec:photo_cleaning}
The photometric quality filtering follows the same prescriptions as in V25 and \cite{Milone2012} with the exception of substituting the outlier rejection with an updated procedure  designed to be robust against crowding, detector edge effects, and magnitude-dependent scatter while preserving the intrinsic width of the cluster sequences. As in V25 and following the same nomenclature,  the outlier rejection is applied  to define the ``well measured" sample, before reddening correction and ``data cuts". 
\begin{figure*}[t]
    \centering
    \includegraphics[width=\textwidth]{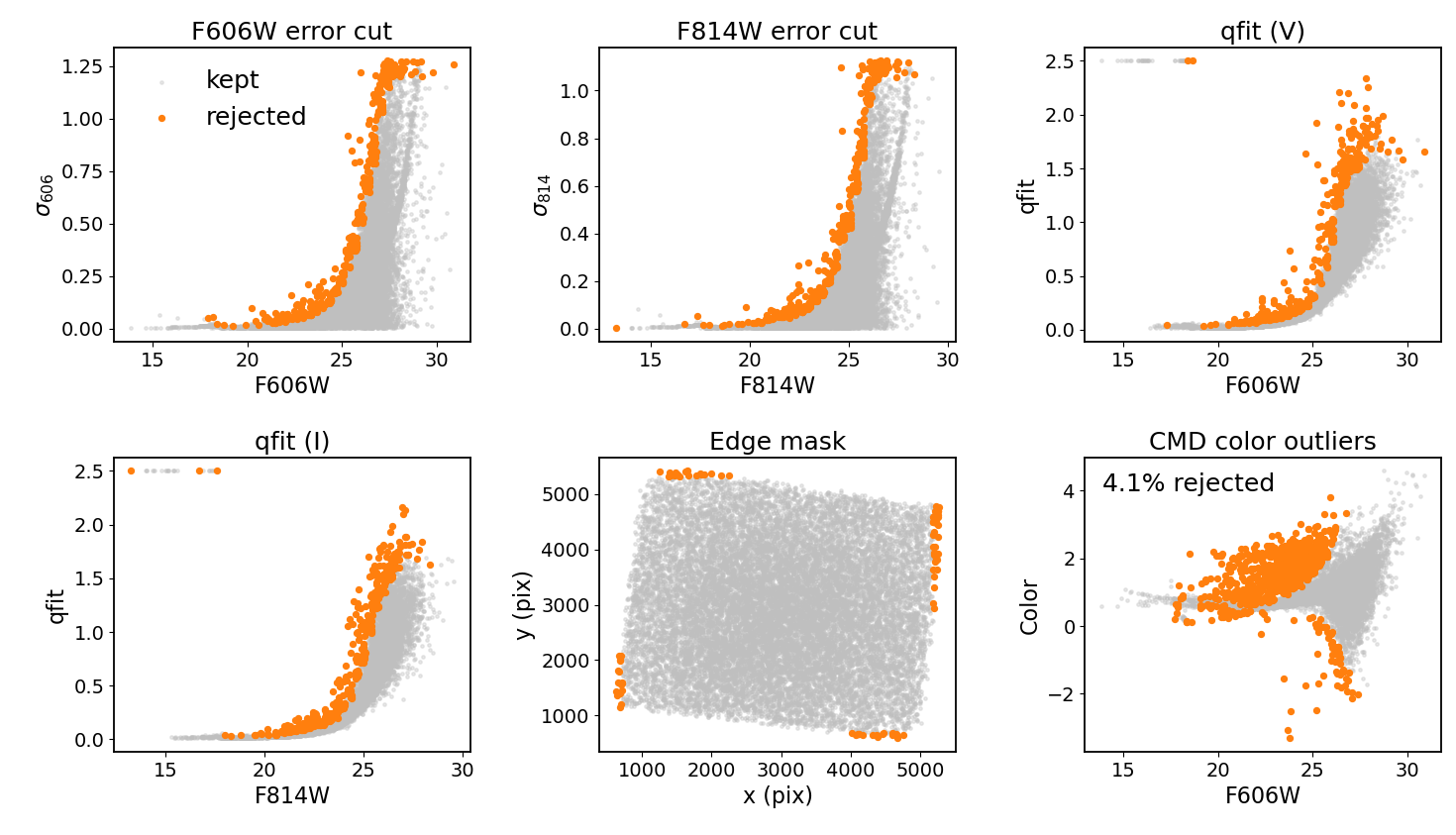}
    \hfill
    \includegraphics[width=\textwidth]{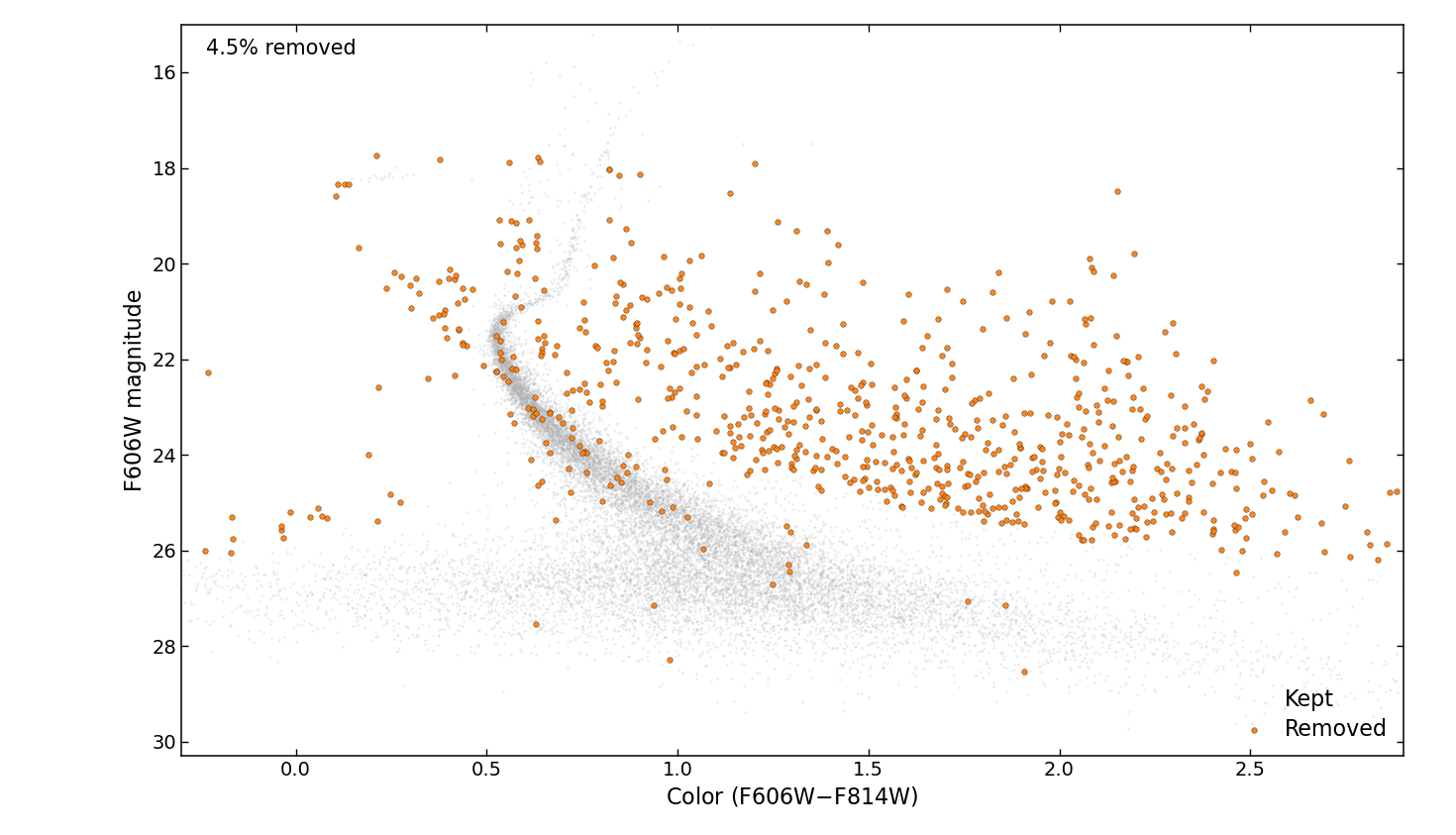}
    \caption{\textbf{Cleaning diagnostics and final Color--Magnitude Diagram (CMD) selection.}
    \emph{Top:} Quality-control projections used to define the cleaning masks, including magnitude--error cuts, \texttt{qfit} thresholds, detector edge masking, and CMD color outlier rejection. Gray points are retained stars and orange points are rejected sources.
    \emph{Bottom:} Final cleaned color--magnitude diagram after all selections, showing that removed stars preferentially occupy field and outlier regions for an representative GC (Arp2). The annotated fraction indicates the small percentage of stars discarded.}
    \label{fig:cleaning}
\end{figure*}
This has three key components.
\paragraph{Edge rejection.}
Stars located close to the detector boundaries are removed. As point spread function (PSF) measurements and background estimates are less reliable near the edges of the field,  we exclude objects that can be affected within a small fractional border of the frame in both spatial directions, defining a spatial mask.

\paragraph{Magnitude–dependent CMD consistency cut.}
The remaining stars are divided into magnitude bins and cleaned directly in color magnitude space.  Within each bin we compute the median color and a robust scatter estimator based on the median absolute deviation.  Stars that deviate from the local stellar locus by more than a few median absolute deviations (including photometric uncertainties) are rejected.  Because this procedure is local in magnitude, it naturally
accounts for the increase of photometric noise at faint magnitudes and avoids the use of global sigma cuts that could bias or artificially narrow the CMD sequences.
\paragraph{Final sample.}
The final ``clean" cluster sample is obtained by combining the spatial and CMD masks.  Importantly, stars rejected by this procedure are retained to construct an empirical model of the field contamination (Section \ref{sec:model_field}).\\

Compared to the previous pipeline, the key change is the adoption of a fully local,
robust CMD-based rejection rather than global thresholds, which improves purity while
preserving genuine sequence broadening expected from multiple stellar populations. The effect of each cleaning stage is illustrated in Fig.~\ref{fig:cleaning}.
Diagnostic projections show the successive rejection of stars with large photometric errors, poor PSF fits, or positions close to the detector edges, followed by the CMD-based color outlier removal.
The final cleaned CMD demonstrates that the rejected sources predominantly populate regions inconsistent with the cluster sequence, while the retained stars define a tight locus.
Overall, only a small fraction of stars ($\sim$4--5\%) is removed, indicating that the procedure primarily suppresses contaminants rather than cluster members.

\subsection{Empirical modeling of field contamination}
\label{sec:model_field}

A further update relative to our previous single population analysis is the explicit construction of an empirical background sample. After the photometric cleaning described in Section ~\ref{sec:photo_cleaning}, the catalog is separated into two subsets: stars that satisfy the cleaning criteria (``kept") and those rejected as outliers
(``removed") which are used to provide an estimate of the field and non member population in color magnitude space.

Using the magnitudes and colors of these rejected stars, we construct a smooth, 
non-parametric probability density via kernel density estimation (KDE).  This provides a data-driven description of the contamination distribution without assuming an analytic model for the field CMD.
This background density is then incorporated into the statistical model described in Section~\ref{weighted_like}, where it forms the field component of the mixture likelihood.

\subsection{Multi-pass ridgeline extraction and duplicate merging}
\label{sec:gmm_passes}

To characterize the cluster sequence in the CMD, we adopt, with a minor improvement,  the same ridgeline extraction framework introduced in our single-population analysis, which forms the backbone of the present procedure. The CMD is divided into magnitude intervals to localize the analysis and, within each bin, the stellar distribution in the two-dimensional CMD space (color and magnitude jointly)
is modeled with a Gaussian mixture. The dominant Gaussian component traces the cluster sequence, providing an estimate of the local ridgeline position together with an associated dispersion and statistical weight. The ensemble of these bin-wise Gaussian mixture fits defines a discrete representation of the cluster locus that is subsequently used in the likelihood evaluation.

In the single-population pipeline this binning and fitting were performed once using a fixed set of magnitude intervals. While effective and adequate  for that application, such a discretization introduces some sensitivity to the arbitrary placement of bin edges, particularly in sparsely populated evolutionary phases where the sequence may straddle adjacent bins.

To mitigate this effect, we extend the procedure here with a multi-pass strategy. The first pass is identical to the binning adopted previously. A second pass repeats the same two-dimensional Gaussian-mixture fitting after shifting the bin grid by approximately half a bin width. This staggered binning improves sampling uniformity and reduces edge-related artifacts by ensuring that stars near bin boundaries in one pass are centrally sampled in the other.

Because the two passes partially overlap, they can produce nearly identical ridgeline points. To avoid double counting, we merge neighboring points from the combined set using a small distance threshold in CMD space. Groups of close points are replaced by a single representative value computed as a weighted average, and their statistical weights are combined accordingly. This merging step yields a unique, consolidated set of ridgeline measurements that is both smooth and insensitive to the initial bin placement.

\begin{figure}[t]
    \centering
    \includegraphics[width=\linewidth]{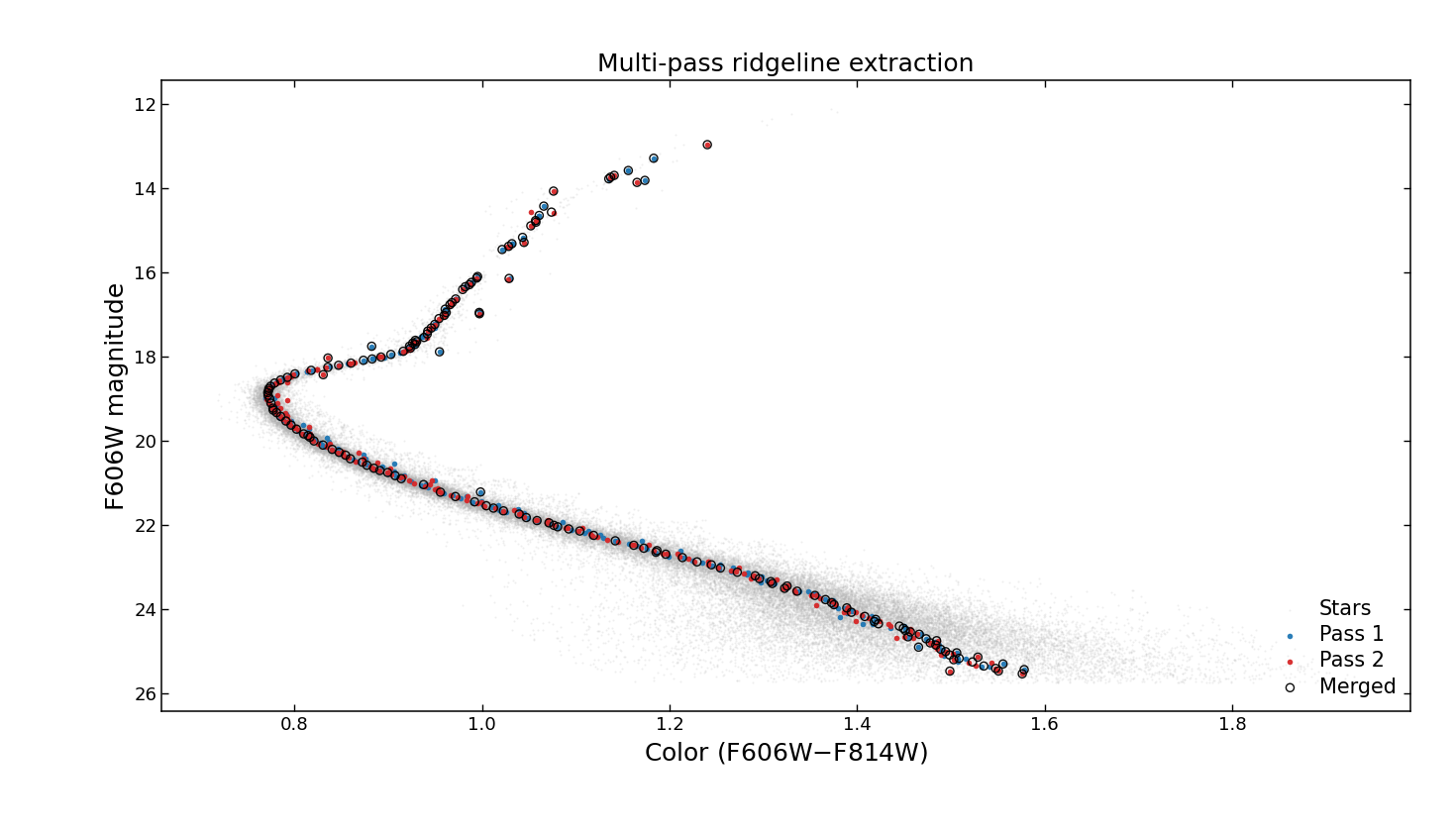}
    \caption{Visualization of the multi-pass ridgeline extraction. Gray points show all cleaned CMD stars. Blue and red markers correspond to the ridgeline estimates obtained from the first and second (half-bin–shifted) passes, respectively. Black points indicate the final merged sequence after duplicate consolidation. The staggered binning improves sampling near bin boundaries and produces a smoother, more robust fiducial line.
    }
    \label{fig:ridgeline_multipass}
\end{figure}

Figure~\ref{fig:ridgeline_multipass} illustrates the outcome of this procedure. The background CMD is shown in gray, while the ridgeline estimates from the first and second passes are displayed separately to highlight the effect of the staggered binning. The final merged sequence follows the high-density locus continuously across the evolutionary phases and provides a stable and well-sampled representation of the cluster fiducial.

Overall, this extension preserves the original two-dimensional Gaussian-mixture framework while providing a more robust and better-sampled description of the cluster sequence, which is particularly important for resolving subtle differences between multiple stellar populations.

\subsection{New parametrization, added free parameters and associated priors}
\label{sec:helium_params}
In the presence of multiple stellar populations, helium abundance becomes a key driver of the morphology of the CMD,
but its inference is strongly degenerate with age and metallicity.
In situations where multiple well-separated sequences are directly observed (for instance when using several filters or ultraviolet photometry),
it can be justified to sample directly in the individual helium mass fractions $(Y_1, Y_2)$ of the two different populations.
In the present case, however, the analysis is based on two optical CMDs only,
for which the populations are not cleanly separated and the observables are primarily sensitive to the \emph{mean} helium content
and to the \emph{relative separation} between populations, rather than to their absolute values independently.
Sampling directly in $(Y_1,Y_2)$ in this regime leads to strong correlations between parameters and poorly constrained directions in parameter space.

For this reason, we adopt a reparametrization in terms of the mean helium abundance $\bar{Y}$, the helium separation, and the population fraction.
Starting from the definition of the population-averaged helium abundance,
\begin{equation}
\bar{Y} = (1-p)\,Y_1 + p\,Y_2,
\end{equation}
where $p$ is the population fraction for population 2,
and defining the helium difference
\begin{equation}
\Delta Y = Y_2 - Y_1,
\end{equation}
we obtain
\begin{equation}
Y_1 = \bar{Y} - p\,\Delta Y, \qquad
Y_2 = \bar{Y} + (1-p)\,\Delta Y.
\end{equation}

This parametrization separates the roles of the helium parameters in a physically meaningful way.
The mean helium abundance $\bar{Y}$ controls the global position of the isochrones and is primarily constrained by the overall locus of the CMD.
The helium difference $\Delta Y$ governs the relative displacement between populations and is constrained by the width and splitting of evolutionary sequences.
The population fraction $p$ controls how strongly each population contributes to the observed CMD and directly modulates the mapping between $\bar{Y}$, $Y_1$, and $Y_2$.

Compared to sampling directly in $(Y_1,Y_2)$, this choice substantially reduces degeneracies and improves sampling efficiency,
while ensuring that both helium abundances remain physically consistent for any allowed combination of parameters.
It also enables the introduction of transparent and physically motivated priors on the mean helium content and on the helium enrichment between populations, rather than on two highly correlated absolute abundances. In practice, the free helium-related parameters of the model are therefore taken to be $(\bar{Y}, \Delta Y, p)$.
The associated priors are described below.

\subsection{Prior on the population fraction}
\label{sec:prior_p}

In addition to the parameters summarized in Table~\ref{tab:priors}, 
the prior on the population fraction $p$ is informed by the spatial distribution of stars within each cluster. It is well established that multiple stellar populations in globular clusters exhibit different radial distributions,
with second-generation stars being more centrally concentrated,
and first-generation stars more spatially extended.
We exploit this information to construct an empirical prior on $p$.

For each cluster, we fit a two-component Gaussian mixture model to the radial distribution of stars.
The component with the smaller characteristic radius is associated with the centrally concentrated population,
while the more extended component is associated with the first population.
From this fit, we estimate the fraction of stars belonging to the extended component,
which we identify with population~1.

We then define a Beta prior on $p$,
\begin{equation}
p \sim \mathrm{Beta}(\alpha, \beta),
\end{equation}
with parameters chosen such that the mean of the distribution matches the inferred fraction of population~1.
Specifically, we set
\begin{equation}
\alpha = \kappa\,\hat{p}, \qquad \beta = \kappa\,(1-\hat{p}),
\end{equation}
where $\hat{p}$ is the fraction obtained from the radial Gaussian-mixture fit,
and $\kappa=5$ controls the concentration of the prior.
This choice provides a weakly informative prior centered on the observed radial segregation,
while allowing the CMD likelihood to dominate the final inference.

\subsection{Weighted Likelihood}
\label{weighted_like}

The introduction of multiple stellar populations increases the complexity of the likelihood
surface and enhances degeneracies between age, helium abundance, and population fractions.
When fitting composite color--magnitude diagrams, different sub-populations can partially
overlap in evolutionary phase, allowing them to compensate each other and making their
separation sensitive to sampling noise and relative star counts.

To address this, we evaluate the likelihood using a richness-weighted formulation and
explicitly model field contamination.  The background probability density is estimated
empirically from the rejected stars using the KDE procedure described in Section~\ref{sec:model_field},
and is incorporated directly as a field component of the likelihood.

For the two-population model, the per-datum likelihood is written as a (Gaussian) mixture of two
stellar populations and a background component,
\begin{equation}
\mathcal{L}_i =
Q \, f \, \mathcal{L}_{1,i} +
Q \, (1-f) \, \mathcal{L}_{2,i} +
(1-Q)\, \mathcal{L}_{{\rm bg},i},
\end{equation}
where $\mathcal{L}_{1,i}$ and $\mathcal{L}_{2,i}$ denote the likelihoods under the two
population models, $\mathcal{L}_{{\rm bg},i}$ is the KDE background term,
$Q$ is the global membership fraction, and $f$ is the relative population fraction.

The total log-likelihood is then constructed as a weighted sum,
\begin{equation}
\ln \mathcal{L}_{\rm tot} = \sum_i p_i \, \ln \mathcal{L}_i,
\end{equation}
where the weights $p_i$ encode the relative number of stars contributing to each CMD
component.  In practice, these weights are derived from the observed star counts in each
CMD panel and normalized such that $\sum_i p_i = 1$, with equal total weight assigned to
each CMD.

This richness-weighted mixture likelihood ensures that the inference is driven by the
statistical support of each CMD component, prevents sparsely populated regions from being
dominated by more numerous ones, and stabilizes the recovery of population fractions and
helium differences.  In the single-population limit, this formulation reduces to the
standard unweighted likelihood adopted in \cite{Valcin25}, ensuring consistency
with our previous analysis.

\subsection{Refined hierarchical analysis}

At the hierarchical level, the global modeling framework remains conceptually similar 
to that adopted in V25, 
in the sense that it aims to infer the intrinsic distribution of globular cluster ages 
from a set of cluster-level posterior constraints. 
However, the quantities entering the hierarchical likelihood are modified 
to reflect the richer structure of the individual age posteriors. 
In V25, the hierarchical analysis was constructed directly from the total stacked age posterior, 
which was well approximated by a Gaussian and could therefore be summarized 
by effective means and variances. 
In the present work, we instead base the hierarchical inputs 
on the Gaussian mixture representations 
which provide a stable and physically meaningful summary of each cluster age distribution, as we detail below.
This change is motivated by the fact that, in the multi-population framework, 
the total stacked posterior is no longer a sufficient statistic. 

At the level of individual clusters, the posterior distributions of the physical parameters are obtained through Bayesian inference using Markov Chain Monte Carlo (MCMC) sampling. For each globular cluster, we explore the joint posterior of the stellar population parameters that determine the morphology of the color--magnitude diagram, including age, metallicity [Fe/H], $\alpha$-enhancement [$\alpha$/Fe], distance, reddening, helium abundance, helium separation between populations, and population fraction.
Sampling is performed using PoCoMC \cite{PocoMC}, a gradient-free MCMC sampler designed to efficiently explore high-dimensional and potentially multi-modal posterior distributions. 
For each cluster, the output of the sampling consists of a set of weighted posterior samples for all parameters, from which we obtain marginalized constraints, correlations, and derived quantities. In particular, we extract the posterior distribution of the cluster age from these samples, which forms the basis for the hierarchical analysis described below.

In V25, cluster age posteriors were well approximated by Gaussian distributions, allowing the hierarchical likelihood to be written directly and analytically in terms of cluster-level means and variances. In the present multi-population framework, however, the inferred age posteriors exhibit significant departures from Gaussianity, including skewness, extended low-age tails, and in some cases weak multimodality. These features 
primarily arise from degeneracies between age and helium abundance when multiple stellar populations are modeled simultaneously.

\begin{figure*}
    \centering
    \includegraphics[width=\textwidth]{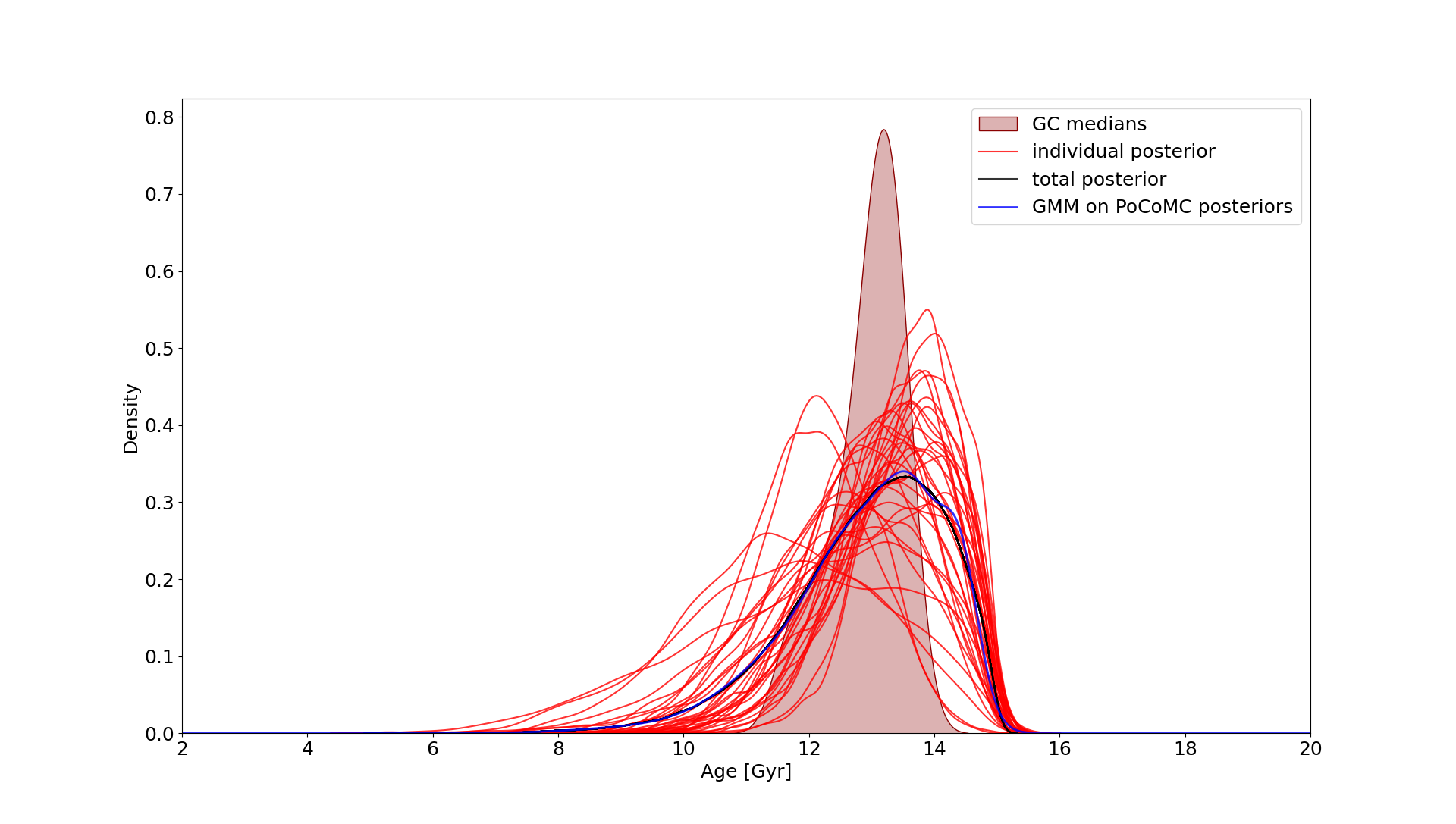}
    \caption{Age posterior distributions for the 36 metal-poor globular clusters ($\mathrm{[Fe/H]} < -1.5$) analyzed in this work. Thin red curves show the individual PoCoMC age posteriors for each cluster. The filled distribution indicates the total posterior obtained by stacking all individual samples, while the black curve represents a kernel density estimate of this combined distribution. The blue curve shows a Gaussian mixture model fit to the aggregated PoCoMC samples. The wide range of posterior shapes, including skewness and extended low-age tails, illustrates the departure from Gaussianity induced by degeneracies between age and helium abundance in the multi-population framework.}
    \label{fig:age_posteriors}
\end{figure*}

This behavior is illustrated in Figure~\ref{fig:age_posteriors}, which shows the individual age posteriors for the metal-poor subsample, along with their combined distribution. The diversity of posterior shapes highlights the limitations of a simple Gaussian approximation and motivates a refinement of the hierarchical modeling strategy. 

Rather than approximating the individual cluster age posteriors  with single Gaussian distributions, we adopt a flexible parametric representation based on Gaussian mixture models (GMMs). For each globular cluster, 
we fit a GMM directly to the PoCoMC age samples, testing models with a range of component numbers (typically between one and a few components), 
and select the preferred representation using the Bayesian and Akaike information criteria (BIC/AIC) to balance goodness of fit and model complexity. This allows the posterior to be described 
as a weighted sum of Gaussian components. This approach provides an efficient and robust approximation of the full posterior shapes, 
capturing skewness, extended tails, and potential weak multimodality 
while retaining a tractable analytical form.

The resulting GMM representations 
are then used as inputs to the hierarchical analysis, 
replacing the single-mean, single-variance summaries 
adopted in V25. The total GMM posterior, obtained by aggregating the individual mixture models, is shown in Figure~\ref{fig:age_posteriors} and is found to provide an excellent approximation to the stacked PoCoMC age distribution. This choice enables the hierarchical model to propagate the full structure of the individual age posteriors, rather than relying on Gaussian approximations that are no longer adequate in the multi-population framework.

As illustrated in Figure~\ref{fig:age_posteriors}, 
the combined distribution is significantly skewed and exhibits extended low-age tails, 
as well as contributions from clusters with weakly multimodal posteriors. 
Relying directly on the total posterior in this regime would bias the hierarchical inference 
toward younger ages. 
This effect is further illustrated by the kernel density estimate constructed from the medians 
of the individual cluster posteriors, which peaks at systematically lower ages, 
reflecting the influence of the asymmetric low-age tails rather than the central mass 
of the cluster-level distributions.
The revised hierarchical construction therefore preserves the original structure of the model, 
while updating the level at which cluster information is summarized, 
ensuring that the hierarchical inference is driven by the full shape of the individual posteriors 
rather than by features of the aggregated distribution alone.

The parameters explored by the sampler and the adopted priors  are as follows (see also table~\ref{tab:priors}). 
At the level of individual clusters, the inference is performed jointly over the stellar population parameters that shape the CMD, including age, metallicity, [$\alpha$/Fe], distance, extinction, Q and the helium parameters $(\bar{Y}, \Delta Y, p)$ introduced in Section~\ref{sec:helium_params}. 
The cluster age and helium parameters are the primary quantities of interest, while distance, extinction, $R_V$, and chemical abundances act as nuisance parameters and are marginalized over in the final constraints.
All parameters are assigned physically motivated priors (see table \ref{tab:priors}). 
Where external measurements are available, we adopt Gaussian priors centered on literature values, in particular for metallicity, [$\alpha$/Fe], and distance. 
For extinction, we relax the fixed uncertainty used in the single-population analysis and instead adopt a magnitude-dependent prior width,
$\delta_{A_V} = 0.02 + 0.1\,A_V$, 
to account for the larger uncertainties affecting highly reddened clusters. 
The population fraction $p$ is assigned a $\mathrm{Beta}(50,5)$ prior, favoring cluster-dominated solutions, while the helium parameters follow the physically motivated priors described in Section~\ref{sec:helium_params}. 
Final constraints on the cluster age are obtained after marginalizing over all nuisance parameters.

\begin{table}
\centering
\begin{tabular}{|c|c|c|}
\hline
Parameter & Uniform prior (bounds) & Additional prior \\
\hline\hline

Age & $U(0,15)$ Gyr & -- \\ \hline

$[\mathrm{Fe/H}]$ & $U(-2.5,0.5)$ dex 
& Gaussian: Ref.~\cite{Harris}, $\sigma=0.2$ dex \\ \hline

Distance & $U(0,\infty)$ 
& Gaussian: Ref.~\cite{Baumgardt} \\ \hline

$[\alpha/\mathrm{Fe}]$ & $U(0,0.4)$ 
& Gaussian: Ref.~\cite{Recio-Blanco}, $\sigma=0.1$ \\ \hline

Extinction & $U(0,3)$ 
& Gaussian: Ref.~\cite{Harris}, $\sigma = 0.02 + 0.1\,A_V$ \\ \hline

$R_V$ & $U(1.5,5)$ & -- \\ \hline

$\bar{Y}$ & $U(0.20,0.40)$ 
& Gaussian: $\sigma = 0.07$\footnotemark \\ \hline

$\Delta Y$ & $U(0,0.15)$ 
& -- \\ \hline

$p$ & $U(0,1)$ 
& Beta$(\alpha,\beta)$ (see Section~\ref{sec:prior_p}) \\ \hline

$Q$ & $U(0,1)$ 
& Beta$(50,5)$ \\ \hline

\end{tabular}
\caption{
Adopted priors on the model parameters. 
All parameters have uniform priors within the indicated bounds. 
Additional Gaussian or Beta priors are included where external information is available 
or to regularize poorly constrained directions of parameter space. 
The first three parameters (age, metallicity, distance) are the main parameters of interest, 
while the remaining parameters are treated as nuisance parameters.
}
\label{tab:priors}
\end{table}

\footnotetext{
We tested a narrower Gaussian prior on $\bar{Y}$ with $\sigma = 0.05$ and found consistent posterior distributions,
indicating that the inference on $\bar{Y}$ is not prior dominated. 
We therefore adopt the more conservative choice $\sigma = 0.07$ in the fiducial analysis.
}

\section{Results}

In order to construct a homogeneous chronometer sample, we restrict our hierarchical analysis to the most metal-poor globular clusters in our HST catalog, selecting systems with $\mathrm{[Fe/H]} < -1.5$. This metallicity cut isolates the oldest Galactic clusters and minimizes residual age--metallicity degeneracies in the isochrone modeling. It results in a subsample of 36 clusters out of the full set of 69 HST clusters, which forms the basis of the hierarchical inference presented below.

A detailed assessment of the impact of priors on the inferred parameters is presented in Appendix~A.
Here we summarize the main conclusions.
The parameters of primary interest, namely the age and metallicity of each cluster,
are well constrained by the CMD data and are not prior-dominated: their posterior distributions are significantly narrower than the adopted priors,
indicating that the constraints are driven by the photometric information.
In contrast, the distance is more weakly constrained by the CMD alone and remains partially prior dominated,
as illustrated in Fig.~8.
This reflects the well-known degeneracy between distance, extinction, and age in optical CMDs,
and motivates the use of external constraints on the distance (i.e.\cite{Baumgardt}).
Among the nuisance parameters, the extinction and $\alpha$-enhancement are partially informed by their Gaussian priors (i.e.\cite{Recio-Blanco, Harris},
while the membership parameter $Q$ and the population fraction $p$ are regularized by Beta priors,
which help stabilize the inference without driving the posterior distributions.
The helium parameters $(\bar{Y}, \Delta Y, p)$ are of particular interest in the context of multiple populations.
We find that the mean helium abundance $\bar{Y}$ is not prior dominated,
as verified by varying the width of its Gaussian prior without affecting the inferred values.
The helium spread $\Delta Y$, for which only weak bounds are imposed,
is primarily constrained by the morphology and width of the CMD sequences.
In addition, the inferred values of $Y_1$ are consistent with primordial helium,
as shown in Fig.~\ref{fig:heliumdist}.
A more detailed discussion of these effects, as well as comparisons with literature constraints,
is provided in Appendix~A.

\subsection{Hierarchical age distribution of metal-poor globular clusters}

\begin{figure}
    \centering
    \includegraphics[width=\linewidth]{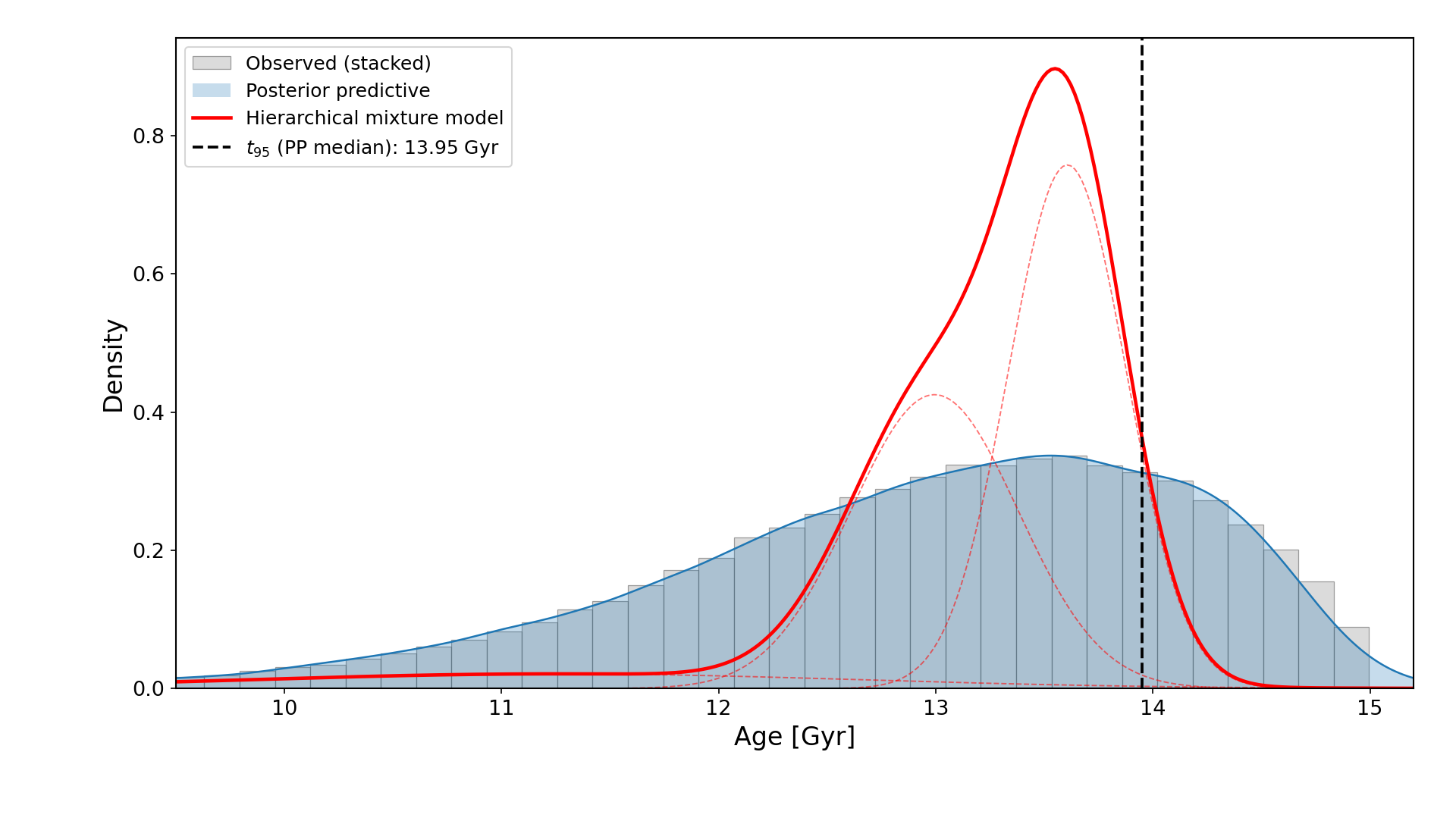}
    \caption{
    Hierarchical reconstruction of the age distribution of metal-poor globular clusters.
    Grey: stacked cluster-level posteriors; blue: posterior predictive distribution;
    red: inferred intrinsic mixture model (dashed lines indicate individual components).
    The three components have posterior mean ages which are Normal distributions characterised by a mean $\mu$ and variance $\sigma$ as:
    $(\mu,\,\sigma) \simeq (11.25,\,1.40)$, $(12.99,\,0.38)$, and $(13.61,\,0.25)$~Gyr.
    The black dashed line marks the posterior predictive 95th percentile, $t_{95}$ (see text for interpretation).
    }
    \label{fig:mixture}
\end{figure}

We apply the hierarchical mixture model to the sample of  36 metal-poor globular clusters
($\mathrm{[Fe/H]} < -1.5$).
Figure~\ref{fig:mixture} shows the resulting reconstruction of the age distribution.
The grey histogram represents the stacked cluster-level age posteriors,
the blue curve the posterior predictive distribution,
and the red curve the inferred intrinsic age distribution described by a Gaussian mixture.
The individual mixture components are shown as dashed lines. The posterior predictive distribution reproduces the observed stacked age distribution well,
including its overall width and its pronounced asymmetry toward younger ages.
In contrast, the intrinsic mixture distribution is narrower and structured,
indicating that a substantial fraction of the apparent spread in the stacked posteriors
is driven by measurement uncertainty and internal cluster degeneracies.
The hierarchical model therefore enables a separation between the observationally broadened age distribution
and the underlying population-level structure.

The inferred intrinsic distribution favors multiple age components,
including a dominant old population with posterior mean
$\mu_{\rm old} = 13.61~\mathrm{Gyr}$ and intrinsic dispersion
$\sigma_{\rm old} = 0.25~\mathrm{Gyr}$.

This oldest component provides a natural definition of the age of the oldest globular clusters in the sample,
as it isolates the dominant population associated with the earliest globular cluster formation epoch,
separating it from both observational broadening and from younger sub-populations
(e.g., accreted clusters).
Although the stellar modeling includes two populations within each cluster,
the Gaussian mixture applied to the \emph{cluster-level} age distribution is not constrained to the same number of components.

The additional components trace populations at younger characteristic ages.
We find two such components, centered at $12.99 \pm 0.33$~Gyr and $11.25 \pm 0.52$~Gyr.
The component at $\sim 13.0$~Gyr is relatively narrow and contains a significant fraction of the clusters,
suggesting a secondary concentration in the age distribution.
This may reflect intrinsic scatter in the early assembly history of the Milky Way,
or indicate the presence of multiple formation phases.
In contrast, the youngest component at $\sim 11.3$~Gyr is broader and subdominant,
indicating a more heterogeneous population of clusters.
This component likely traces systems with more diverse formation histories,
including clusters formed or accreted at later times.

The presence of three components therefore reflects the global structure of the age distribution across clusters,
rather than the number of stellar populations within individual clusters.
In particular, the additional component at younger ages can be understood as providing flexibility
to describe a non-Gaussian, skewed age distribution,
rather than necessarily indicating a distinct physical population.

\subsection{Extreme-age characterization}
\begin{figure}
    \centering
    \includegraphics[width=\linewidth]{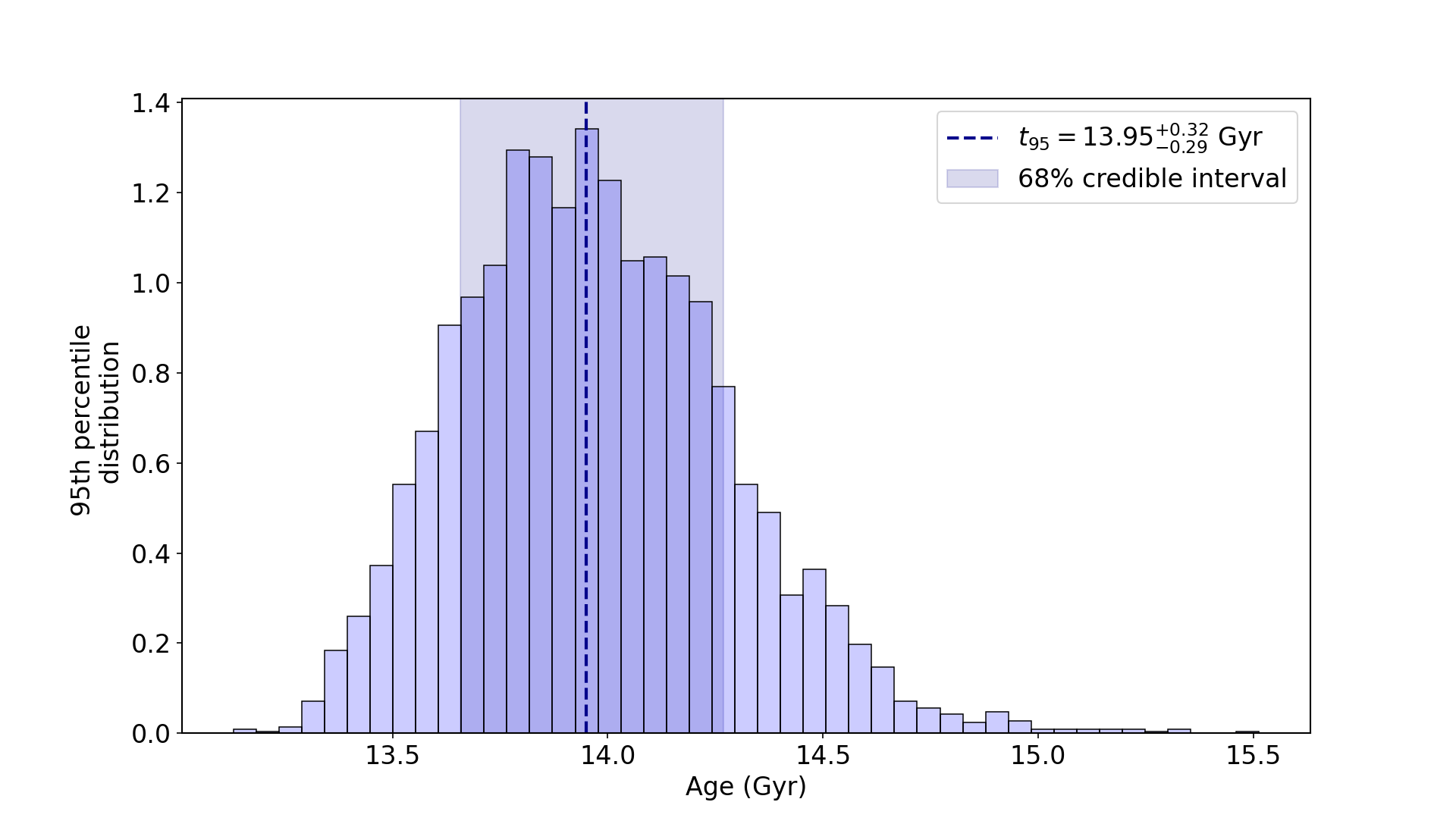}
    \caption{
    Posterior distribution of the 95th percentile $t_{95}$ of the globular cluster age distribution,
    derived from the hierarchical posterior predictive model.
    The dashed line indicates the median, and the shaded region the central 68\% credible interval.
    }
    \label{fig:t95}
\end{figure}

In addition to the parameters of the intrinsic mixture components,
we characterize the extreme-age behavior of the reconstructed globular cluster population
using the 95th percentile of the posterior predictive age distribution, $t_{95}$.
The location of this percentile is indicated by the dashed black line in Figure~\ref{fig:mixture},
and its posterior distribution is shown in Figure~\ref{fig:t95}. From this distribution, we infer
\begin{equation}
t_{95} = 13.95^{+0.32}_{-0.29}~\mathrm{Gyr},
\end{equation}
where the quoted values correspond to the median and central 68\% credible interval. This statistic provides a complementary characterization of the old-age tail of the distribution, and is sensitive to both the location and intrinsic width of the oldest population as well as to uncertainties in the hierarchical model parameters.

\subsection{Age of the Universe}
The estimates obtained from the hierarchical mixture model correspond to present-day globular cluster ages.
To convert them into an estimate of the age of the Universe, we account for the delay between the Big Bang and the formation
of the first globular clusters. We parametrize this delay by $\Delta t$, which captures the onset of star formation,
early chemical enrichment, and the assembly of the first bound stellar systems. Both theoretical work on primordial star formation and recent observations of the high-redshift Universe
indicate that the emergence of the first long-lived stellar systems occurred very early in cosmic history.
Models of Population~III and early Population~II star formation predict that the first sites of star formation
appear at redshifts well above $z \sim 10$, but not at arbitrarily early times, with an onset regulated by the build-up
of the first collapsed structures.
In parallel, spectroscopic studies of galaxies at $z \gtrsim 10$ now directly confirm the presence of star-forming systems
within the first few hundred million years after the Big Bang. As in V20 and V25, we do not adopt a fixed delay, but use the full probability distribution for $\Delta t$ obtained following Ref.~\cite{Jimenez2019}, marginalizing over $H_0$, $\Omega_{m,0}$, and the formation redshift $z_f$.
Taken together, these results place the formation of the first bound, metal-poor stellar populations
within roughly a few $\times 10^8$ years after the Big Bang, corresponding to a characteristic delay
$\Delta t \simeq 0.2$--$0.4$~Gyr.
A recent synthesis by Tomasetti et al.\ (2025, Section~4.1) reviews these constraints and adopts the same physically motivated range.

Following this literature, we write
\begin{equation}
t_U = t_{\rm GC} + \Delta t,
\end{equation}
where $t_{\rm GC}$ denotes the age of the oldest globular clusters.
In the present framework we define $t_{\rm GC}$ as the mean of the oldest intrinsic age component inferred from the hierarchical model,
$t_{\rm GC} \equiv \mu_{\rm old}$, which isolates the dominant population associated with the earliest globular cluster formation epoch,
separating it from both observational broadening and from younger sub-populations (e.g., accreted clusters).

From the hierarchical mixture analysis we obtain
\begin{equation}
t_{\rm GC} = \mu_{\rm old} = 13.61 \pm 0.25~\mathrm{Gyr},
\end{equation}
where the quoted uncertainty is statistical.

In this work we adopt a conservative reference value of $\Delta t = 0.2$~Gyr,
corresponding to the earliest plausible onset of globular cluster formation within the physically motivated range.
Combining $t_{\rm GC}$ with this formation delay, we infer an age of the Universe of
\begin{equation}
t_U = 13.81 \pm 0.25 ({\rm stat.})~\mathrm{Gyr},
\end{equation}
where the quoted uncertainty is statistical.
Systematic uncertainties associated with stellar modeling, distances, reddening, and the adopted formation delay
are discussed in V21 and account for an additional $\sigma_{\rm sys.}=0.23$ (we refer the reader to V1 for more details).

In addition to this best-estimate determination, we also propagate the extreme-age characterization based on the posterior predictive distribution.
Defining $t_{U,95} = t_{95} + \Delta t$, and using
$t_{95} = 13.95^{+0.32}_{-0.29}$~Gyr, we obtain
\begin{equation}
t_{U,95} = 14.15^{+0.32}_{-0.29} (\rm stat.)~\mathrm{Gyr},
\end{equation}
which provides a complementary description of the extent of the old-age tail of the reconstructed globular cluster population.

Both estimates are fully consistent with each other and with independent cosmological age determinations.
In particular, they are compatible with the $\Lambda$CDM cosmic age inferred from the CMB,
$t_U \simeq 13.8~\mathrm{Gyr}$, while providing an astrophysics-based determination derived from Galactic GCs.

\section{Conclusions and Outlook}

In this work, we have reassessed the ages of 69 Galactic globular clusters by explicitly modeling two independent stellar populations with distinct ages, metallicities, and helium abundances. Our analysis demonstrates that allowing for multiple populations introduces only minimal changes to the inferred ages. As shown in Fig.~\ref{fig:mixture}, the ages of both stellar components remain statistically indistinguishable from each other and are in excellent agreement with the single-population results of V25, with differences well within $0.5\,\sigma$ of the statistical error (Fig.~\ref{fig:V25andthis}). This confirms the robustness of globular clusters-based cosmic chronometry against uncertainties in stellar population complexity.

The two-component Gaussian mixture modelling successfully recovers the helium spread $  \Delta Y  $ between populations (Fig.~\ref{fig:heliumave}) and first-population fractions $  f_{\rm P1}  $ (Fig.~\ref{fig:popratio}), in good agreement with prior studies \citep{Milone17,Milone18}. Furthermore, Fig.~\ref{fig:heliumdist} shows that the helium abundance differences between the two populations are small but clearly measurable, consistent with independent determinations in the literature. Astrophysical validation (Appendix~\ref{sec:amr}) further reinforces these findings: the reconstructed age-metallicity relation recovers the classic bifurcation between in-situ and accreted branches, with progenitor-specific relations closely matching established expectations and supporting distinct chemical enrichment histories. Well-known correlations of multiple populations   quantities ($  \overline{Y}  $, $  \Delta Y  $, and $  f_{\rm P1}  $) with cluster mass and central concentration are also preserved (Appendix~\ref{sec:correlations}). This overall consistency in helium content, metallicity trends, and population parameters confirms the reliability of our methodology in disentangling subtle chemical signatures while robustly preserving age constraints.

Taken together, these results reinforce the use of globular clusters as precise and largely model-independent probes of cosmic time. The inferred Universe age of  $13.81\pm 0.25({\rm stat.})\pm 0.23 ({\rm sys.})\,\mathrm{Gyr}$ remains compatible with $\Lambda$CDM predictions tied to early-Universe measurements, while continuing to provide a valuable independent check on late-time determinations that favor a higher $H_0$. The persistence of this agreement, even when introducing additional freedom in the stellar population modeling, strengthens the case for GCs as a powerful complementary probe of cosmology. Future work incorporating JWST observations of high-redshift clusters and more sophisticated population synthesis models will further refine these constraints providing additional perspective  into the 
current tensions in cosmic expansion measurements.

\begin{acknowledgments}
Funding for the work of RJ and LV  was partially provided by
project PID2022-141125NB-I00, and the “Center of Excellence Maria de Maeztu 2025-2029” award to the ICCUB funded by grant CEX2024-001451-M funded by MICI-
U/AEI/10.13039/501100011033. Based on data products from observations made with ESO Telescopes at the La Silla Paranal Observatory under ESO programme ID179.A-2005 and on data products produced by TERAPIX and the Cambridge Astronomy Survey Unit on behalf  of the UltraVISTA consortium.
\end{acknowledgments}

%\bibliography{sources}
\bibliographystyle{JHEP}

\providecommand{\href}[2]{#2}\begingroup\raggedright\endgroup

\appendix 
\section{Pipeline validation and cluster--level results}
\label{sec:pipeline_results}

In this section we present the cluster--level results obtained with the analysis pipeline
described in Section~2, focusing on the recovery of stellar population parameters, internal
consistency checks, and comparison with previous single--population analyses. Figures~\ref{fig:posterior_distrib},
\ref{fig:heliumdist}, \ref{fig:heliumave}, \ref{fig:V25andthis} and  \ref{fig:popratio}
summarize the performance of the pipeline across the full sample of 69 Galactic globular
clusters.

\subsection{Recovered parameters and prior sensitivity}

\begin{figure}[h]
    \centering
    \includegraphics[width=0.49\linewidth]{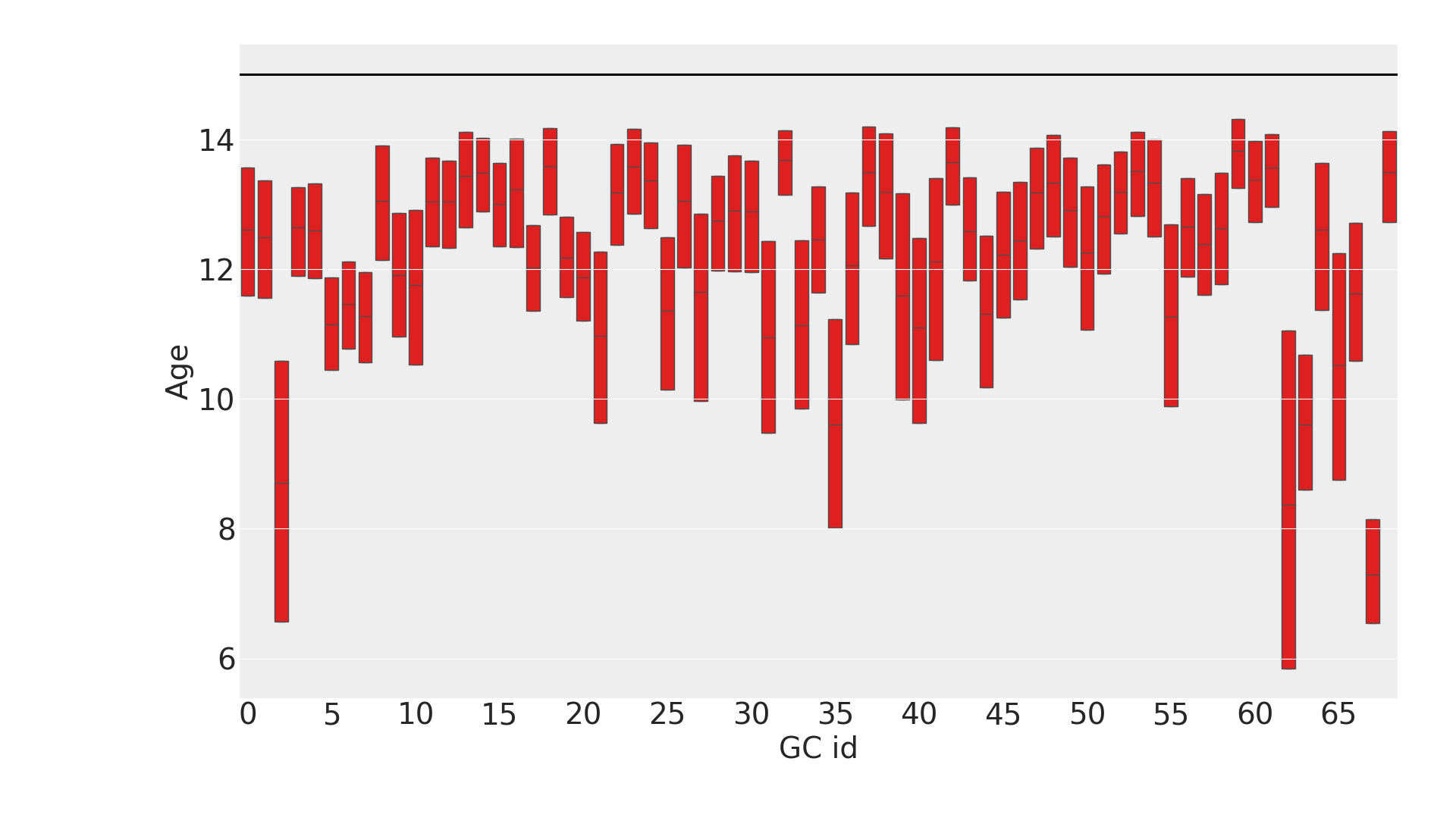}
    \includegraphics[width=0.49\linewidth]{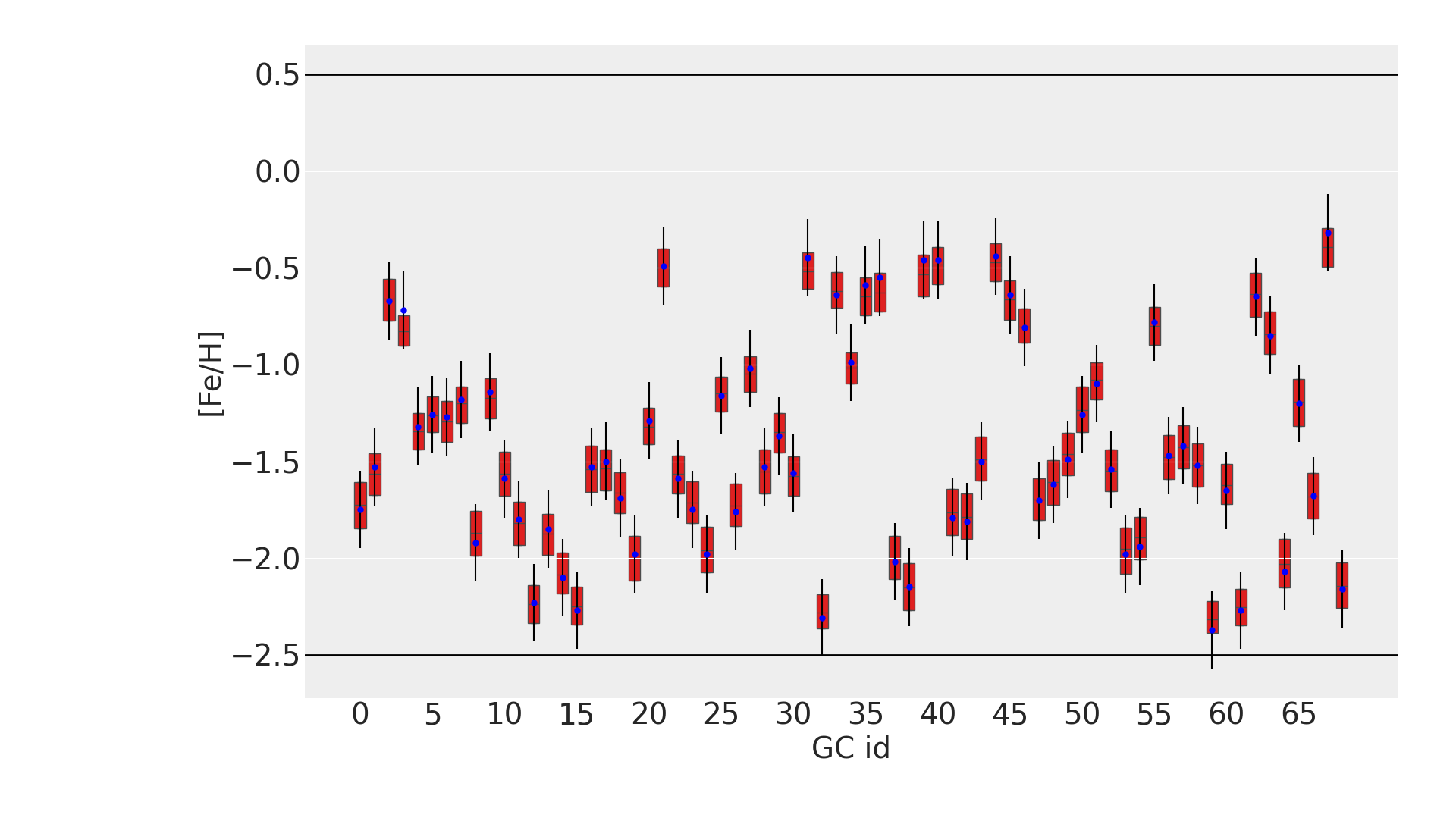}
    \includegraphics[width=0.49\linewidth]{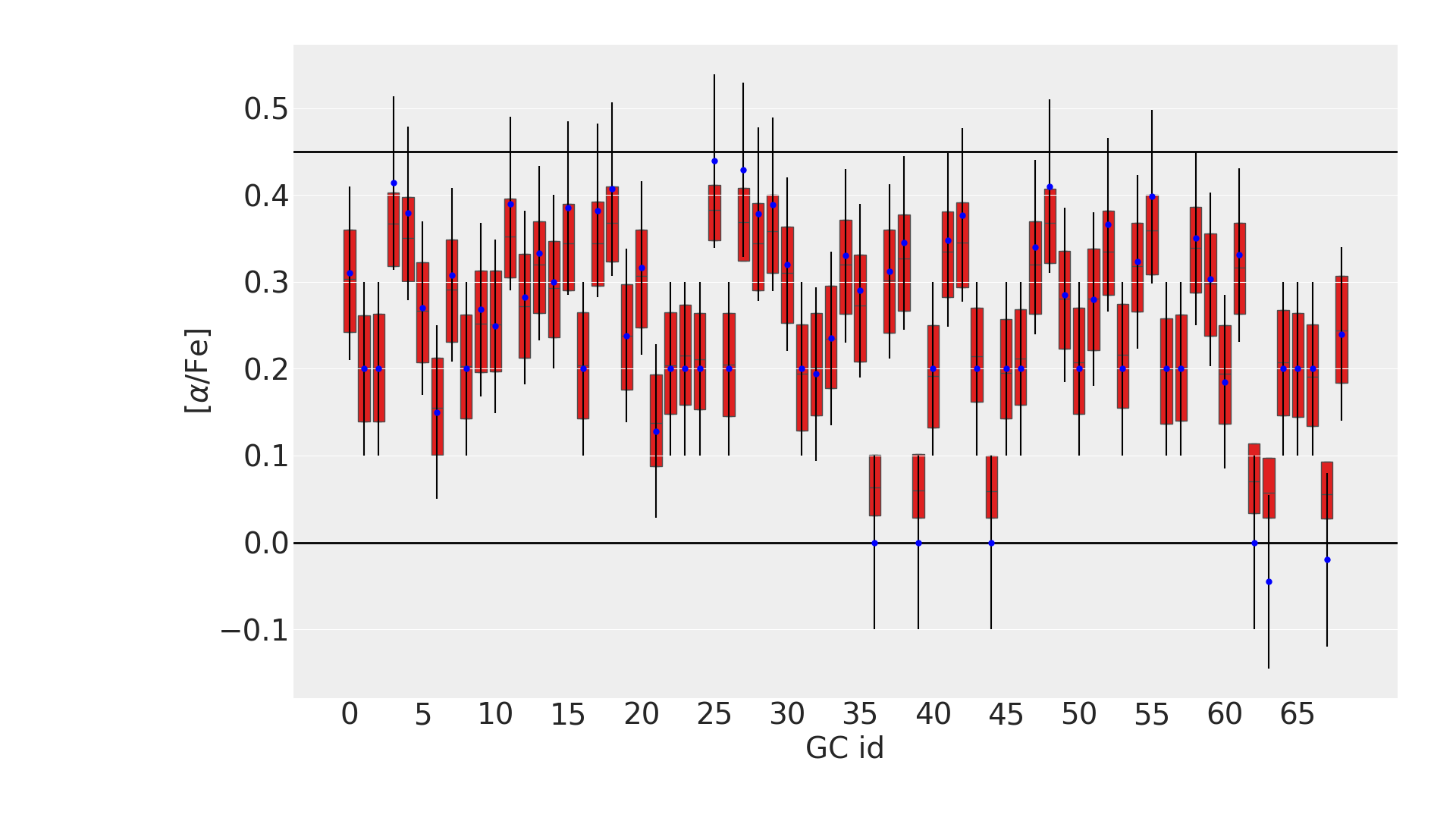}
    \includegraphics[width=0.49\linewidth]{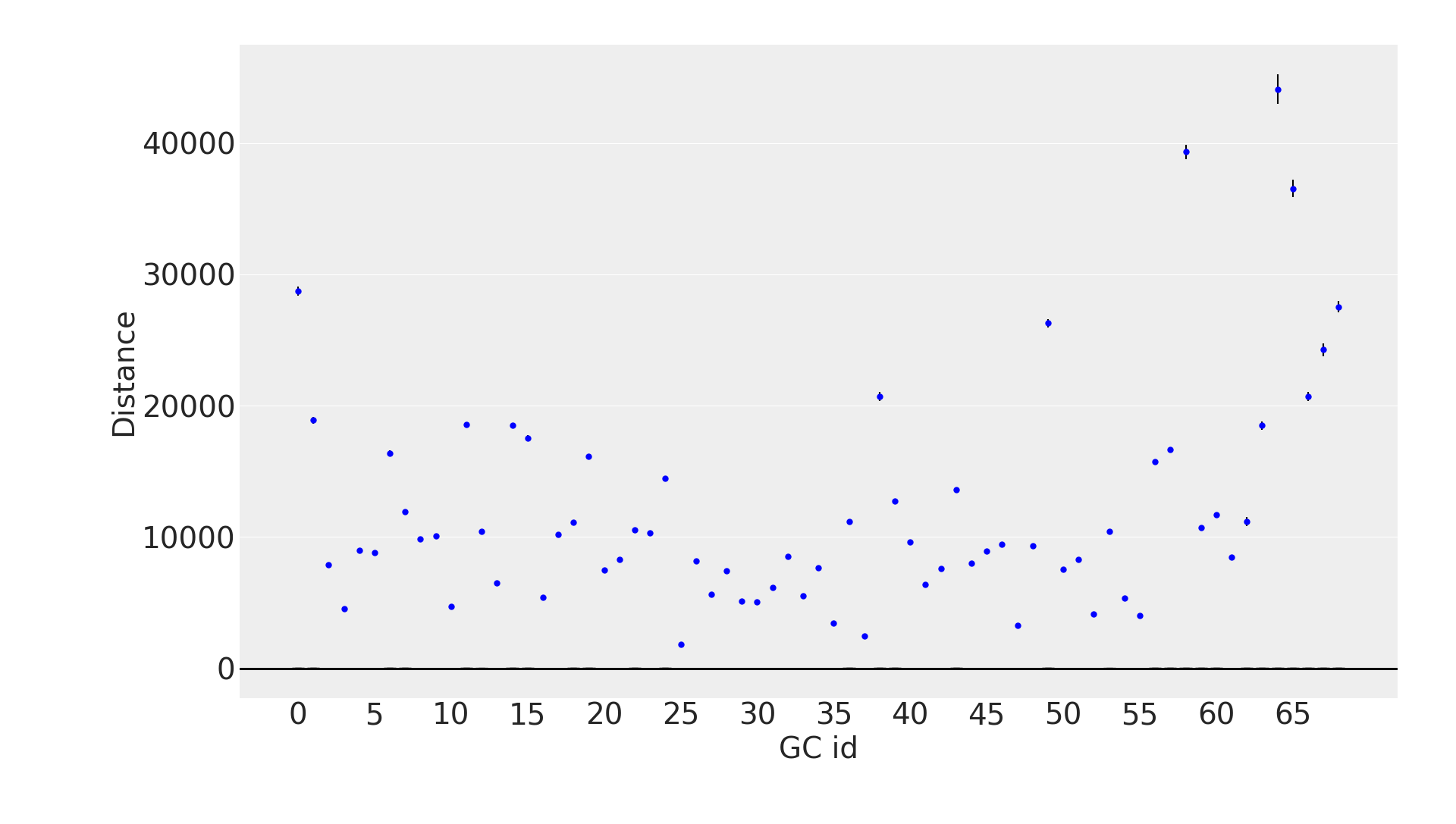}
    \includegraphics[width=0.49\linewidth]{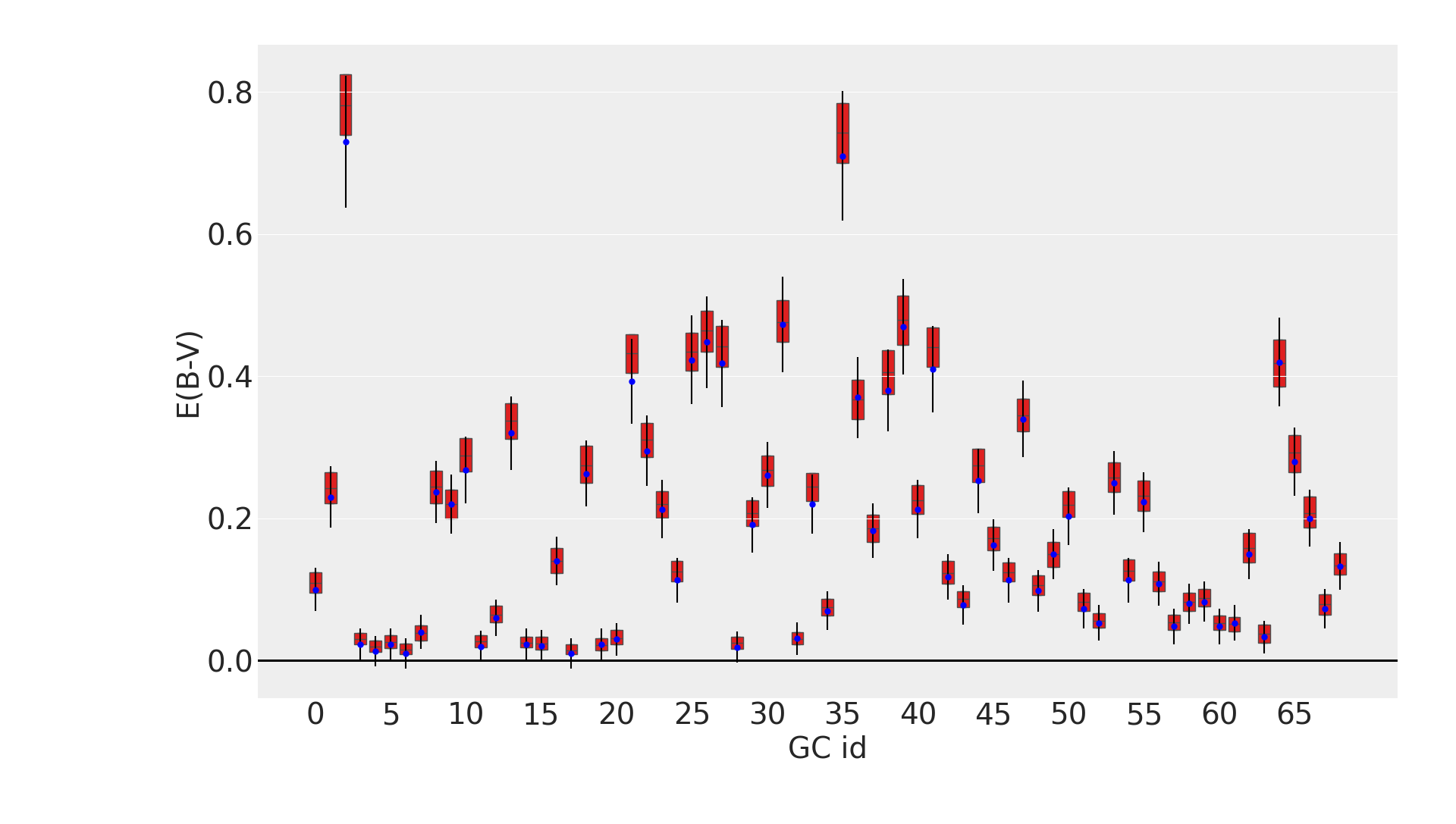}
    \includegraphics[width=0.49\linewidth]{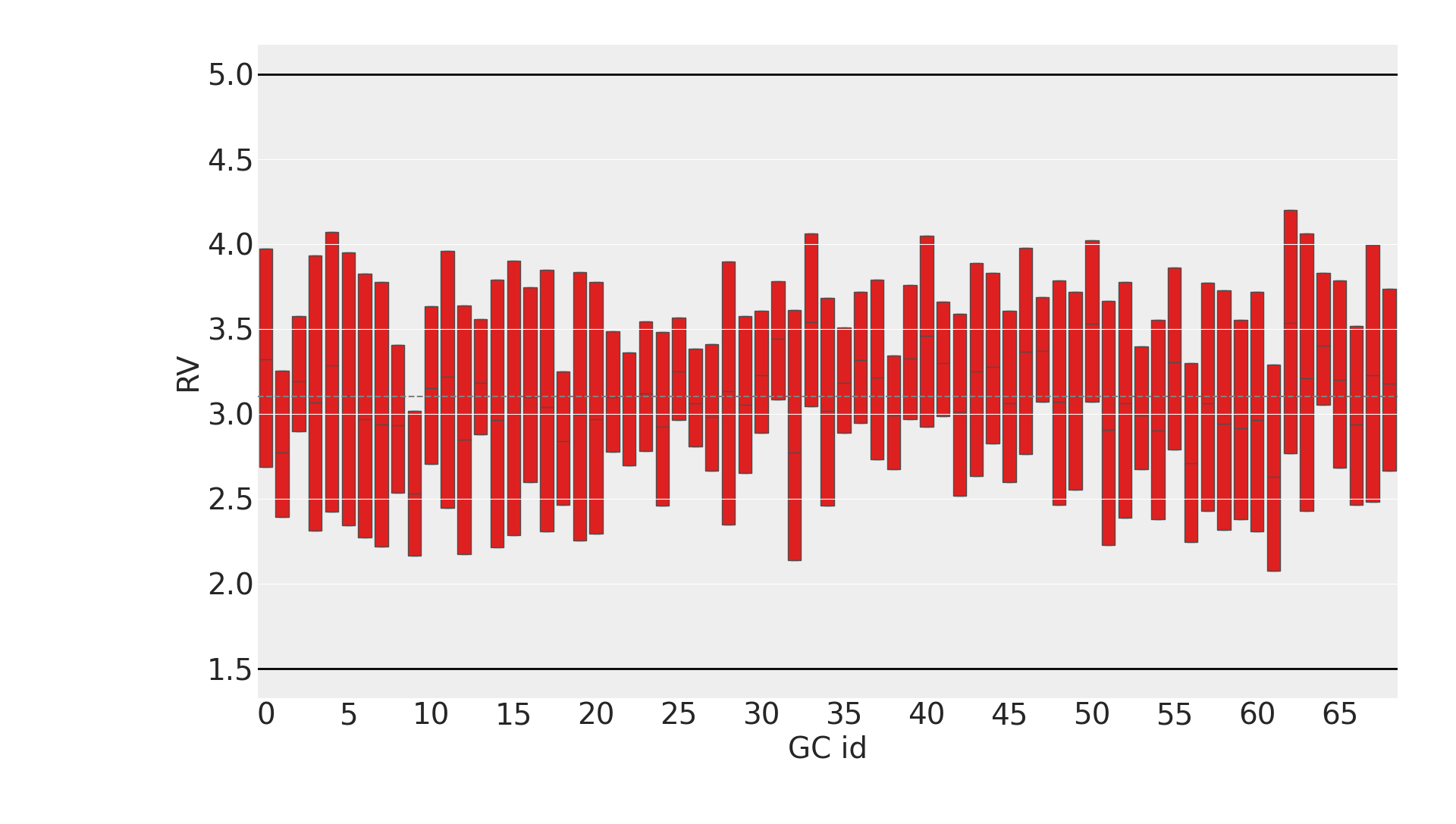}
    \includegraphics[width=0.49\linewidth]{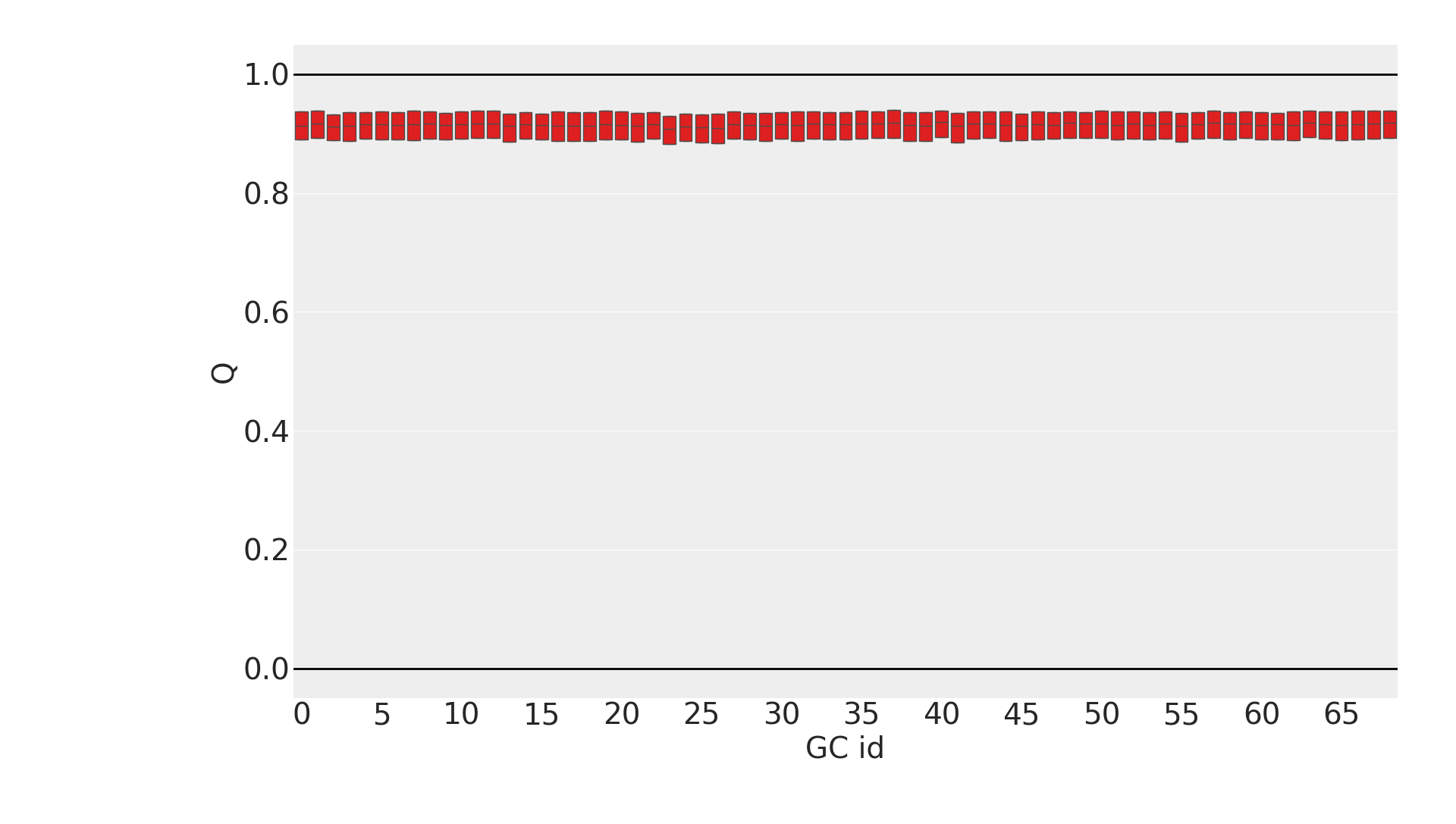}
    \includegraphics[width=0.49\linewidth]{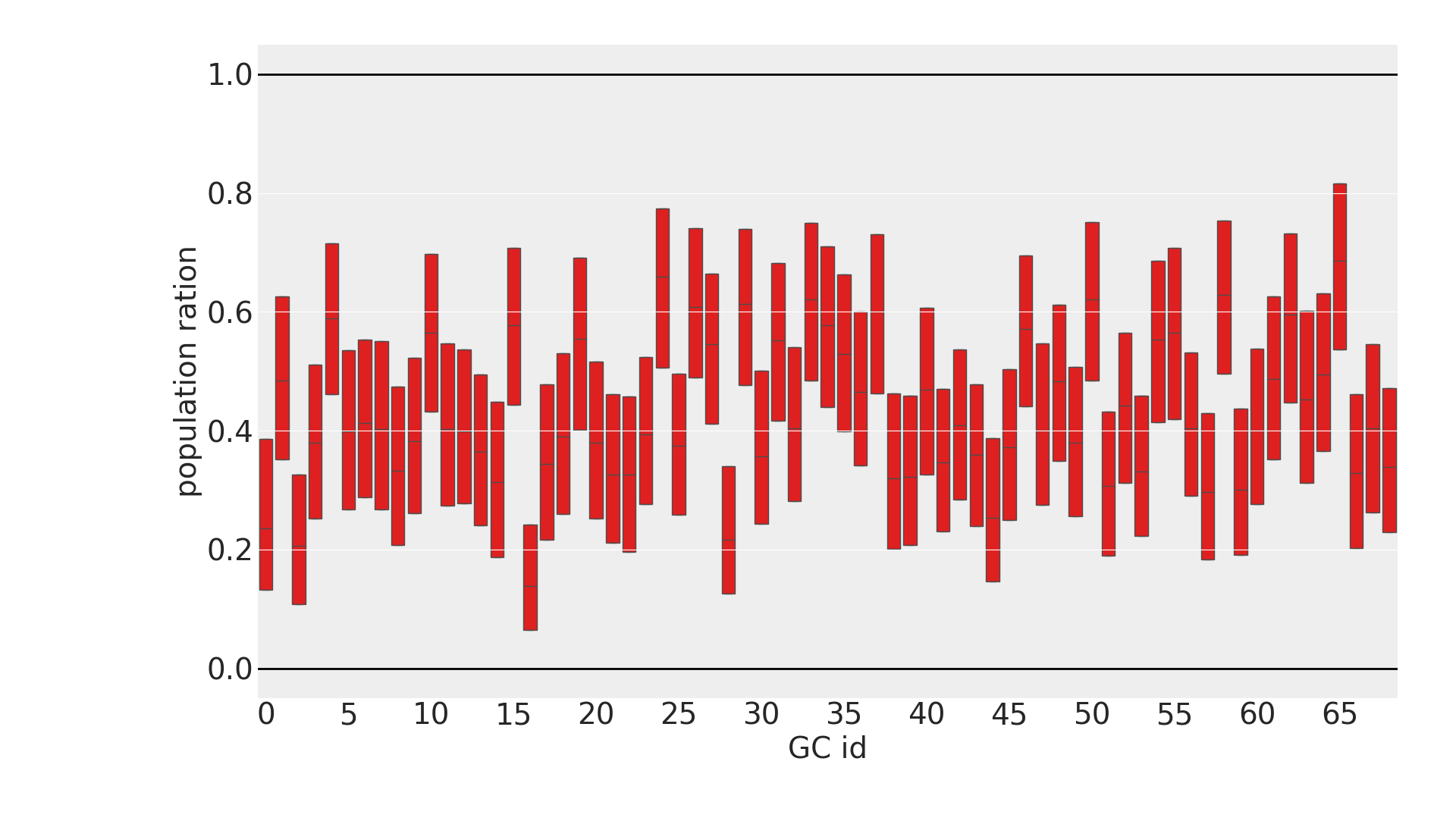}
    \includegraphics[width=0.49\linewidth]{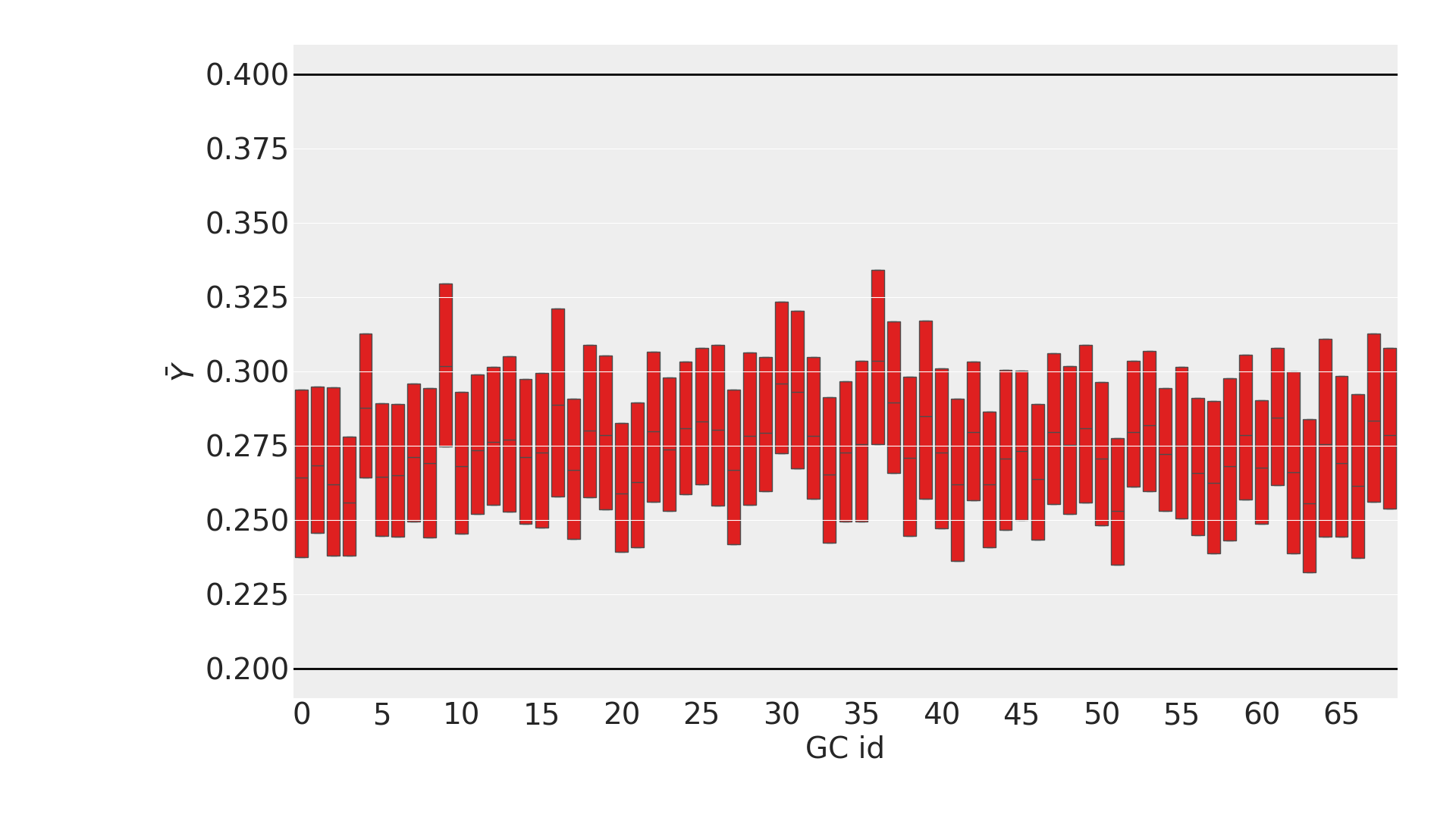}
    \includegraphics[width=0.49\linewidth]{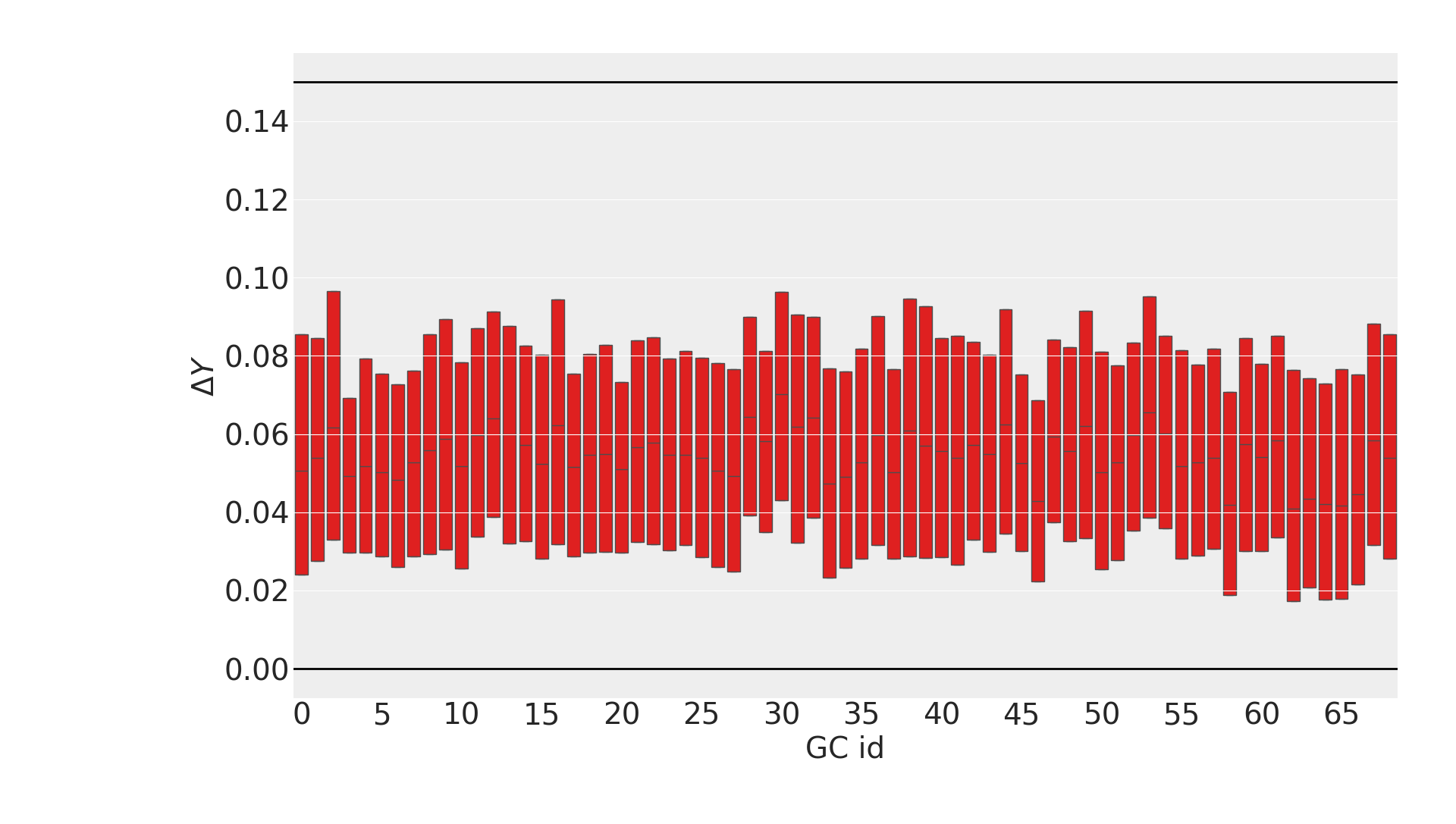}
   
    \caption{Marginalized best fit values for each of the 69 GC in the sample, with corresponding 1D uncertainty bounds (as red bars) and the priors used for each parameter, horizontal thick black  lines are the hard priors, thin vertical lines are the Gaussian priors (reported only where applied). The distance determinations are dominated by the prior, see discussion in~\cite{Valcin25}.}   \label{fig:posterior_distrib}
\end{figure}

Figure~\ref{fig:posterior_distrib} displays the marginalized posterior constraints for the main model parameters of
each GC, including age, metallicity, distance modulus, reddening, mean helium
abundance, helium separation between populations, and population fractions. The corresponding
priors are shown for reference.

\FloatBarrier

As expected, parameters that are weakly constrained by the optical color--magnitude
diagrams alone---notably distance and reddening---remain largely prior-dominated, consistent
with analysis presented in \cite{Valcin2021,Valcin25}. 
In contrast, ages, metallicities, and helium--related parameters
are significantly constrained by the data, demonstrating that the full CMD morphology
contains sufficient information to disentangle these quantities even in the presence of
multiple stellar populations.

Importantly, no pathological behavior or unphysical parameter combinations are observed
across the sample, indicating that the reparametrization adopted for helium abundance and
population fractions leads to stable and well--behaved posteriors.

\subsection{Helium abundances and internal consistency}

\begin{figure}
    \centering
    \includegraphics[width=\linewidth]{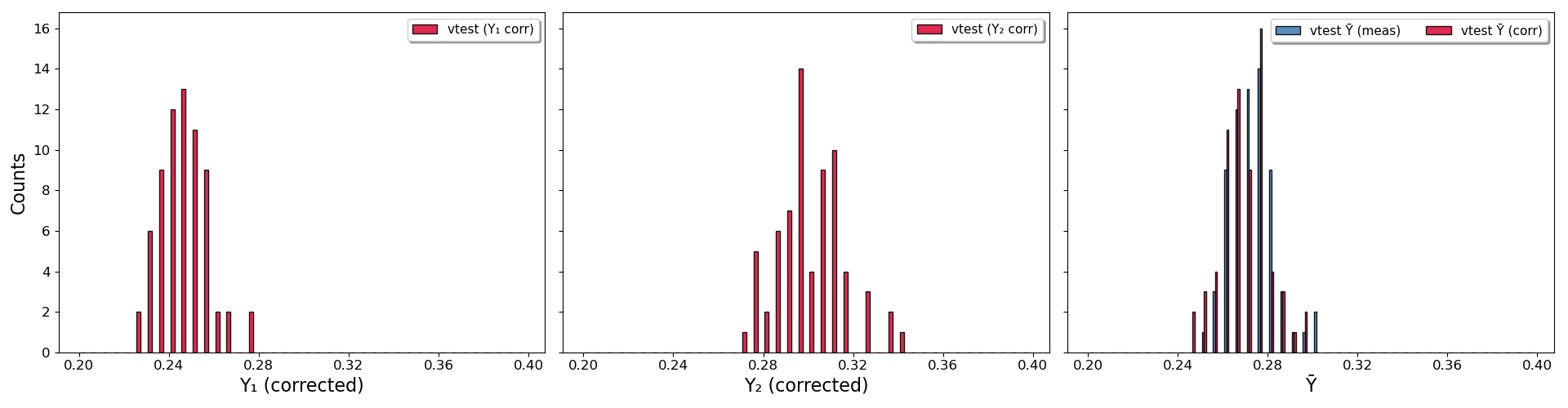}
    \caption{Helium abundance distributions for the cluster sample. 
    Left and middle panels show the inferred distributions of $Y_1$ and $Y_2$, computed from the posterior constraints on the mean helium abundance $\bar{Y}$ and helium separation $\Delta Y$. The right panel compares the measured (uncorrected) and corrected mean helium values. The recovered $Y_1$ values are consistent with primordial helium within uncertainties.}
    \label{fig:heliumdist}
\end{figure}

\begin{figure}
    \centering
    \includegraphics[width=0.8\linewidth]{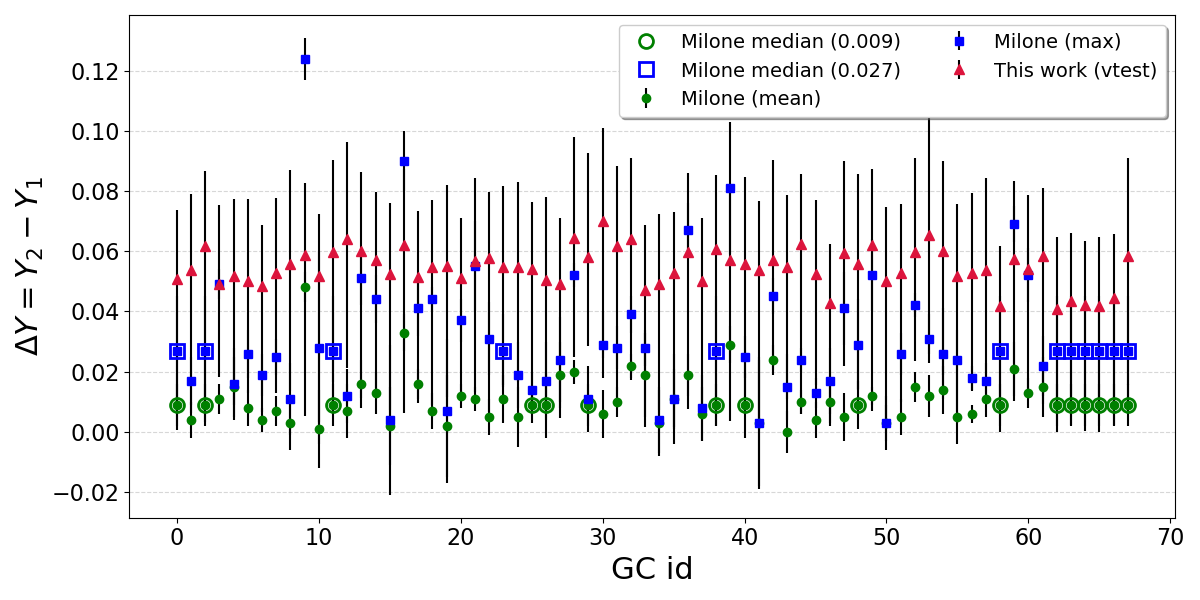}
    \caption{Helium difference $\Delta Y = Y_2 - Y_1$ for the cluster sample. 
Red triangles denote the values inferred in this work, with credible intervals from the posterior distribution. 
Blue and green symbols show literature estimates from Milone et al. 
The overall level and scatter of helium enrichment are consistent.}
    \label{fig:heliumave}
\end{figure}

Figures~\ref{fig:heliumdist} and~\ref{fig:heliumave} present the distributions of the inferred helium abundances for the two stellar populations, as well as the helium difference $\Delta Y$ between them. The typical
helium separations are small but non--zero, and fall comfortably within the range reported in previous
studies.

A detailed comparison of our $  \Delta Y  $ values with those from \cite{Milone18} (shown in Figure~\ref{fig:heliumave}) reveals a small but systematic positive offset. In the subsample of 51 clusters with reliable measurements of both mean and maximum $  \Delta Y  $, our mean $  \Delta Y  $ is higher by $0.045 \pm 0.009$,dex on average than the value of Ref.~\cite{Milone18}  (median offset $  +0.046  $,dex, RMS scatter $0.009$,dex). The maximum $  \Delta Y  $ is higher by $0.024 \pm 0.021$,dex (median $  +0.029  $,dex, RMS $0.021$,dex).

We examined possible correlations between these offsets ($  \Delta Y_{\rm this~work} - \Delta Y_{\rm M18}  $) and key cluster parameters. The most notable (though modest) dependence is with metallicity [Fe/H], where the offsets exhibit a moderately weak negative correlation: Pearson $  r \approx -0.32  $ ($  p = 0.022  $) and Spearman $  \rho \approx -0.36  $ ($  p = 0.009  $) for the mean $  \Delta Y  $. This indicates that our $  \Delta Y  $ values tend to be relatively larger (compared to \cite{Milone18}) in more metal-poor clusters. A marginal positive correlation with cluster age is also present ($  r \approx +0.26  $, $  p \approx 0.06  $--$0.09$  ), but it remains statistically weak. No significant trends emerge with [  $\alpha$/Fe] or $  E(B-V)  $ (all $  |r| \lesssim 0.12  $, $  p > 0.4  $). Weighted linear fits yield very shallow slopes in all cases ($  |\mathrm{slope}| \lesssim 0.006  $,dex/dex for [Fe/H], and substantially smaller otherwise), showing that any parameter dependence is subtle and does not account for the dominant systematic offset of $  \sim 0.02  $--$0.045$,dex.

The cluster-to-cluster scatter in $  \Delta Y  $ remains moderate, with no evidence for extreme He enhancements that could significantly bias age determinations. These findings support the central assumption of our analysis pipeline: although helium abundance is partially degenerate with age, the multi-band photometric constraints are sufficiently constraining to prevent large systematic errors in the inferred cluster ages.

\FloatBarrier

\subsection{Comparison with single--population ages}

\begin{figure}
    \centering
    \includegraphics[width=0.8\linewidth]{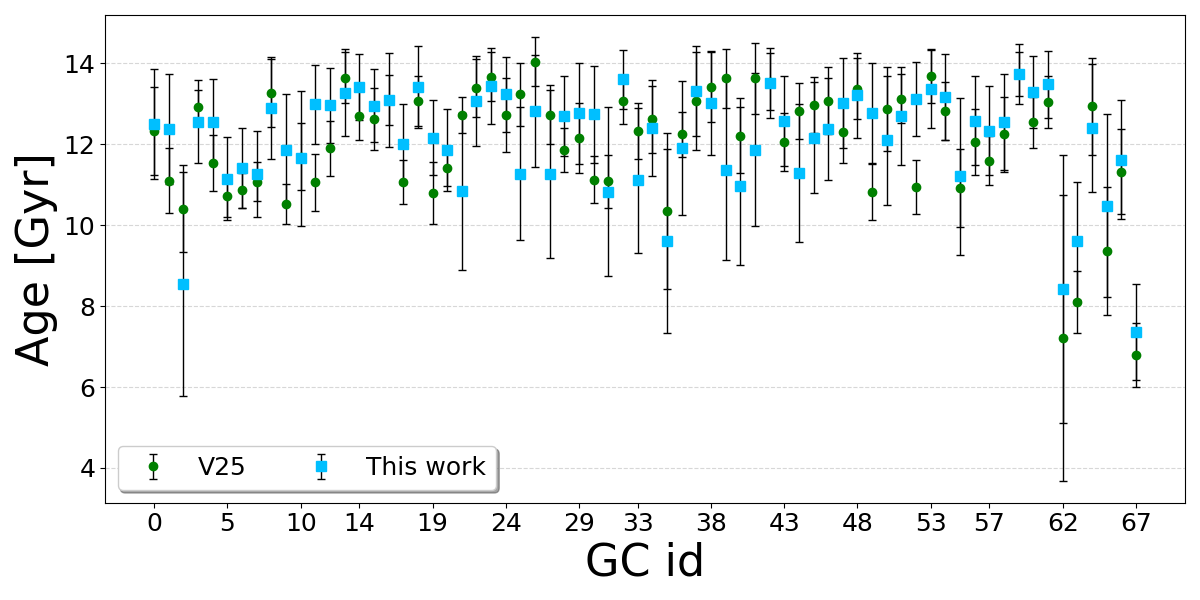}
    \caption{Comparison between V25 ages and the ages of the two populations from this study. As expected, the ages from the two populations are similar to each other and perfectly consistent with those derived in V25.}
    \label{fig:V25andthis}
\end{figure}

A direct comparison between the ages obtained in this work and those from the previous
single--population analysis \cite{Valcin25} is shown in Figure~\ref{fig:V25andthis}. The ages inferred for the two populations within each cluster are statistically indistinguishable from each other and from the
single--population ages.

This result demonstrates that introducing additional freedom to model MPs does not lead to significant changes in the inferred GC ages.
Differences are well within $0.6\sigma$ for the vast majority of clusters, confirming that
the age estimates are robust against assumptions about internal population complexity. We find no evidence for a systematic shift toward either younger or older ages relative to V25. The differences are cluster-by-cluster and consistent with statistical fluctuations within the uncertainties.

\FloatBarrier

\subsection{Population fractions and age independence}
\label{sec:MP_fractions}

\begin{figure}
    \centering
    \includegraphics[width=0.8\linewidth]{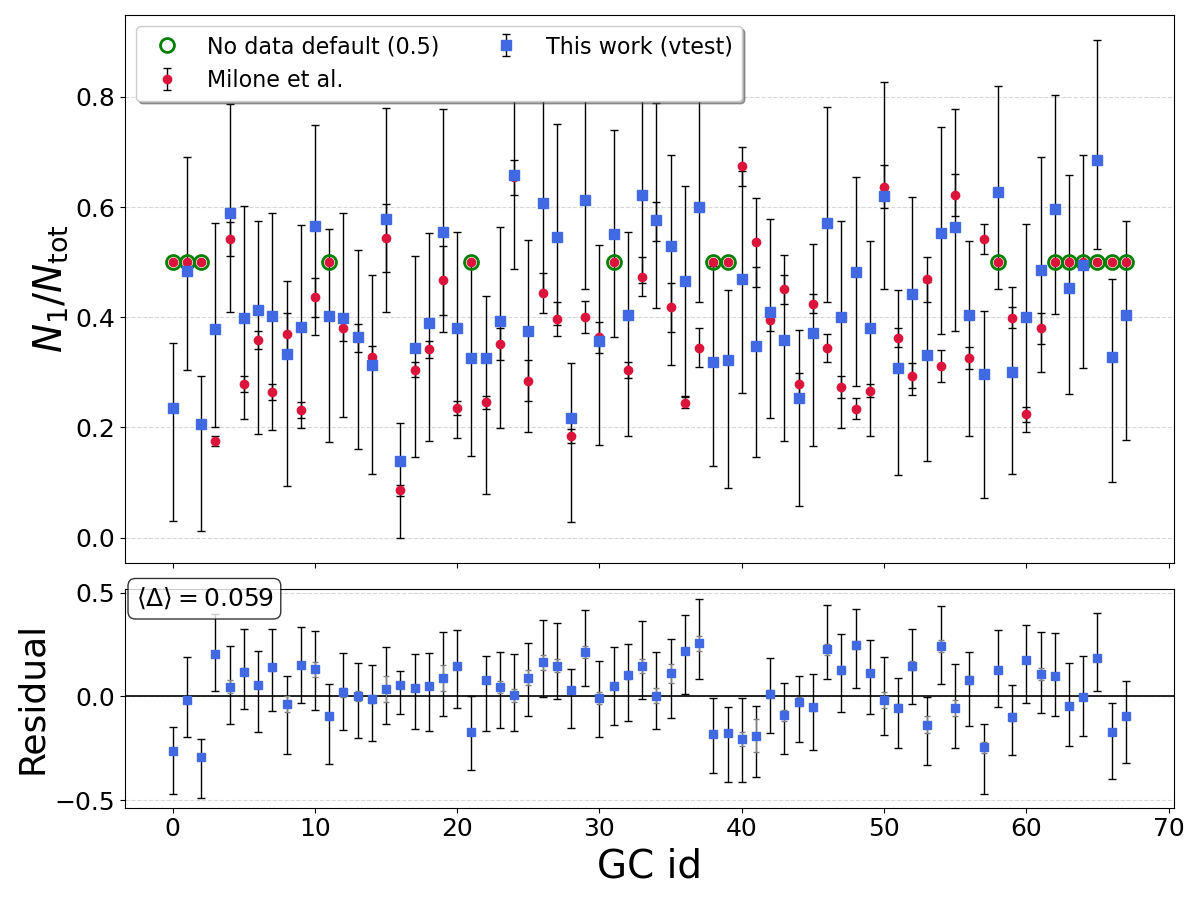}
    \caption{First-population fractions $f_{\rm P1} = N_1/N_{\rm tot}$ for the 69 clusters. Blue squares show the values inferred in this work (credible intervals shown), red circles denote literature measurements from Milone et al.~\cite{Milone17}, and green symbols indicate clusters without published Milone values. The lower panel shows residuals.}
    \label{fig:popratio}
\end{figure}

Figure~\ref{fig:popratio} compares the inferred first-population fractions $  f_{\rm P1}  $ from this work with literature values from \cite{Milone17}. Our $  f_{\rm P1}  $ estimates exhibit a small but systematic positive offset relative to those reported by \cite{Milone17}, with a mean difference $  \Delta f_{\rm P1} \approx +0.053  $ (median $  \approx +0.047  $) across the 53 \footnote{The remaining 16 clusters do not have published Milone \cite{Milone17} values and are therefore excluded from this comparison.}  clusters with reliable measurements in both studies. This indicates that our analysis consistently assigns a slightly higher proportion of first-population stars. The scatter around this offset is substantial (RMS $  \approx 0.118  $) and significantly exceeds the level expected from formal measurement uncertainties alone. Consequently, the two sets of $  f_{\rm P1}  $ estimates are not compatible within their quoted errors.

A significant positive correlation exists between our derived $  f_{\rm P1}  $ values and the residuals $  \Delta f_{\rm P1}  $ (Pearson $  r \approx +0.42  $, $  p \approx 0.002  $, slope $  \approx +0.42 \pm 0.13  $), implying that the systematic overestimation becomes progressively larger as the first-population fraction increases. By contrast, no meaningful dependence of the offset or scatter is detected on cluster age or on the helium abundance difference $  \Delta Y  $. These null trends indicate that the systematic discrepancy is not driven by formation epoch or by the degree of He enhancement.

The most likely origin of this offset and excess scatter is methodological. The study presented in \cite{Milone17} derives population fractions directly from chromosome maps-pseudo-two-colour diagrams that exploit segregation along multiple photometric indices to achieve high-contrast separation of the two main populations with different C, N, O, and He abundances \citep{Piotto15}. In contrast, the present analysis infers $  f_{\rm P1}  $ indirectly through full forward-modelling of the F606W-F814W colour-magnitude diagram morphology within a hierarchical Bayesian framework. These fundamental differences between direct chromosome-map segregation and global CMD fitting provide the most plausible explanation for the observed systematic offset and inflated scatter, rather than any underlying astrophysical dependence on age or helium content. 

Together, Figures~\ref{fig:posterior_distrib}--\ref{fig:popratio} demonstrate that the pipeline delivers stable, physically plausible,
and internally consistent results across the full globular cluster sample. The agreement
with previous work and with independent determinations of population fractions and helium
content provides strong validation of the methodology, and justifies its use in the
hierarchical age analysis presented in the following sections.

\section{Astrophysical Validation of Derived Cluster Properties}
\label{app:astro_val}

To confirm the reliability and astrophysical significance of our homogeneous derivations, we compare the ages and metallicities obtained in this work with independent observational constraints and established trends from the literature. In particular, we test whether we can recover the well-known age-metallicity bifurcation observed for Galactic GCs and derive individual age-metallicity relations for in situ and accreted structures (Sec.~\ref{sec:amr}). Furthermore, we examine whether the multiple poulations (MP)  properties, mean helium abundance $  \overline{Y}  $, helium spread $  \Delta Y  $, and first-population fraction $  f_{\rm P1}  $—reproduce the well-established correlations known from previous studies of MPs in massive, old clusters (Sec.~\ref{sec:correlations}).

\subsection{Age--Metallicity Relations and Progenitor Signatures of the Accreted Milky Way}
\label{sec:amr}

The age-metallicity relation (AMR) of Galactic GCs remains one of the most reliable tracers of their formation environments and the hierarchical assembly history of the Milky Way. A defining feature of this relation is its well-established bifurcation: a steep, relatively metal-rich branch dominated by in-situ clusters formed within the proto-Milky Way, and a shallower, more metal-poor branch primarily populated by accreted clusters originating from low-mass satellite galaxies \citep{Forbes10,Leaman13}.

\begin{figure*}
    \centering
    \includegraphics[width=\linewidth]{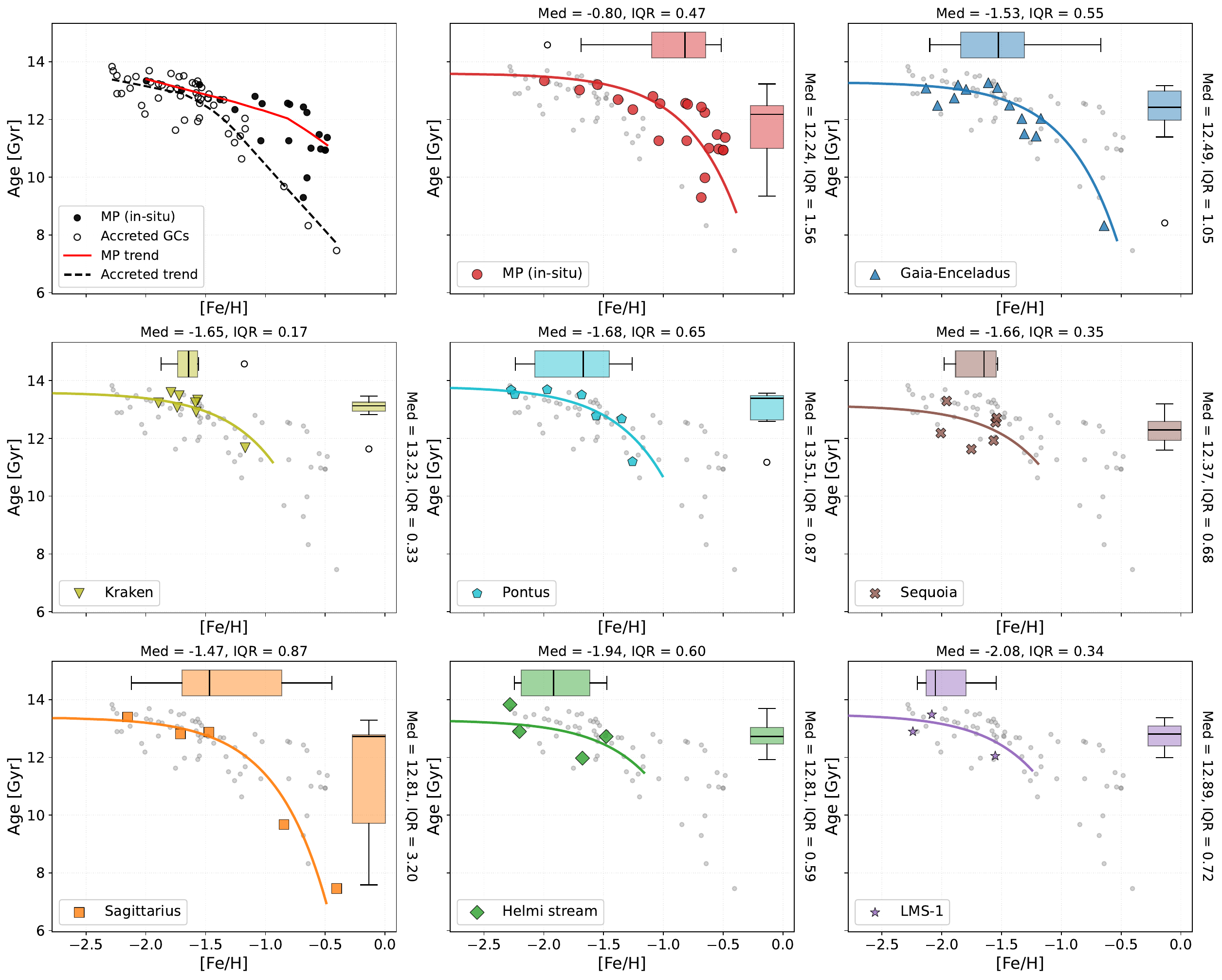}
    \caption{AMrs for the 69 Galactic GCs analyzed in this work, derived homogeneously using our isochrone-fitting framework. The top-left panel displays the overall bifurcated AMR, with trends overlaid for the in-situ (solid red line) and accreted (dashed black line) components to guide the eye. Subsequent panels highlight individual accreted progenitors.
    Clusters are color-coded and symbolized according to their likely progenitor origins, following classifications from the literature (e.g., \cite{Massari19,Massari23,Malhan22,Forbes20,Horta20}).
    Gray background points indicate all clusters not highlighted in a given panel, for reference.
    In each panel, a leaky-box chemical evolution model is superimposed as a solid colored line for illustration. Inset boxplots summarize the median [Fe/H] and age, along with their IQR, for each group.}
\label{fig:age-metallicity-pro}
\end{figure*}

The top-right panel of Figure~\ref{fig:age-metallicity-pro} reconstructs this bifurcated AMR using the homogeneous ages and metallicities derived in the present study, with smoothing trends overlaid for the in-situ and accreted components. This reconstruction closely aligns with established literature trends \citep{Forbes10,Leaman13}, thereby validating the reliability of our age and metallicity derivations and enhancing their astrophysical significance for probing Galactic formation history.

Building on this foundation, we exploit these homogeneous parameters to examine how the AMR varies across distinct progenitor systems. Figure~\ref{fig:age-metallicity-pro} presents individual panels for the in-situ main progenitor (MP), Gaia-Enceladus/Sausage \citep{Helmi18}, Pontus \citep{Malhan22}, Sagittarius \citep{Ibata94}, Sequoia \citep{Myeong19}, the Helmi streams progenitor \citep{Helmi99}, LMS-1 \citep{Yuan20}, and the low-energy (Kraken/Koala-like) group \citep{Massari19,Kruijssen19,Kruijssen20,Forbes20}. Gray background points show all remaining clusters for context, while inset boxplots summarize the median and interquartile range (IQR) in [Fe/H] and age for each group.
Classifications are based primarily on dynamical associations from \cite{Massari19,Massari23}. For recently identified structures, we adopt those from \cite{Malhan22} for the Pontus merger (distinguished from Gaia-Enceladus by distinct dynamical properties) and LMS-1; from \cite{Horta20} for NGC~6535 (chemical assignment); and from \cite{Forbes20} for Palomar~1.
A pronounced age-metallicity anti-correlation is evident across all groups, in agreement with previous studies employing homogeneous single-population age scales \citep{Forbes10,Leaman13,Valcin2020,Valcin2021}. Specifically, the most metal-poor clusters ([Fe/H] $  \lesssim -1.8  $) are preferentially the oldest, whereas metal-richer objects display greater scatter and a systematic shift toward younger ages. For illustrative purposes, each group's AMR panel includes a simple leaky-box chemical evolution model, following the parametrization adopted by \cite{Massari19} and \cite{Forbes20} (see also \cite{Prantzos08,Leaman13}).

Again Our findings align with the main conclusions of previous studies \citep{Massari19,Forbes20,Kruijssen20,Malhan22}. Below, we summarize the key characteristics of each major progenitor, as revealed by the homogeneous ages and metallicities in Fig.~\ref{fig:age-metallicity-pro}:
\begin{itemize}
\item \textbf{In-situ main progenitor (MP; $  N=23  $)}: Defines the highest-normalization AMR branch, with a model [Fe/H] $  \approx -1.26 \pm 0.10  $ dex at 13 Gyr. It has a median [Fe/H] = $  -0.80  $ dex and the youngest median age (12.24 Gyr, IQR = 1.63 Gyr), indicating rapid and extended chemical enrichment in the proto-Milky Way disc/bulge, likely driven by high star-formation efficiency and strong gas retention.
\item \textbf{Gaia-Enceladus/Sausage ($  N=13  $)}: Shows a coherent AMR offset by $\sim$0.6 dex below the MP branch. Clusters span [Fe/H] $  \approx -2.0  $ to $  -0.5  $ dex (median $  -1.53  $ dex). The broad metallicity spread and scaling $  N_{\rm GC} \propto M_{\rm halo}  $ support a relatively massive accreted galaxy with \citep{Helmi18,Malhan22}. 
\item \textbf{Pontus ($  N=7  $)}: Has a 13 Gyr normalization of $  -1.50 \pm 0.15  $ dex, comparable to the low-energy group and higher than most minor accretions. Its metallicity distribution function ranges from [Fe/H] $  \approx -2.3  $ to $  -1.3  $ dex (median $  -1.7  $ dex). Identified via clustering in action-energy space with Gaia EDR3 \citep{Malhan22}, Pontus appears distinct, although some dynamical analyses suggest possible overlap with Gaia-Enceladus when bar effects are considered \citep{Tkachenko23}. Further chemo-dynamical constraints are needed to confirm its independence.
\item \textbf{Sagittarius ($  N=5  $)}: Exhibits the largest age spread (IQR = 3.20 Gyr), consistent with extended star formation and enrichment. The 13 Gyr normalization is relatively high ([Fe/H] = $  -1.59 \pm 0.48  $ dex), with the AMR evolving to higher metallicities at later epochs due to sustained gas retention. This suggests a relatively massive, luminous satellite ($  M_{\rm tot} > 10^9 M_{\odot}  $; \cite{Deason19}), with enrichment aided by prolonged tidal stripping rather than rapid quenching.
\item \textbf{Low-energy (Kraken/Koala-like) group ($  N=9  $)}: Displays remarkably tight distributions in age (median 13.23 Gyr, IQR = 0.33 Gyr) and [Fe/H] (median $  -1.65  $ dex, IQR = 0.17 dex), with normalization similar to Sagittarius. The high inferred yields point to an early, moderately massive accretion \citep{Forbes20,Kruijssen20}.
\item \textbf{Smaller accreted groups} (Helmi streams progenitor, $  N=4  $; Sequoia, $  N=6  $; LMS-1): Exhibit systematically lower AMR normalizations, consistent with lower progenitor masses and shorter star-formation durations.
\end{itemize}

The diversity in AMR slopes, normalizations, and intrinsic scatters across progenitor systems strongly supports the interpretation that the Galactic GC population preserves a fossil record of chemical evolution timescales, star-formation efficiencies, and total masses of their host galaxies or substructures. These trends remain robust within our homogeneous sample, highlighting the critical role of precise, uniform age determinations in reconstructing the Milky Way's hierarchical accretion history.

Importantly, incorporating multiple stellar populations into our isochrone-fitting procedure does not erase or qualitatively alter the AMR. The overall slope, normalization, and scatter align closely with those from our previous analyses using homogeneous single-population age scales \citep{Valcin2020,Valcin2021,Valcin25}. This demonstrates that the Galactic GC AMR is shaped primarily by the formation epoch and chemical enrichment history of the progenitor systems, rather than by internal population complexity within individual clusters.

\begin{figure}
    \centering
    \includegraphics[width=\linewidth]{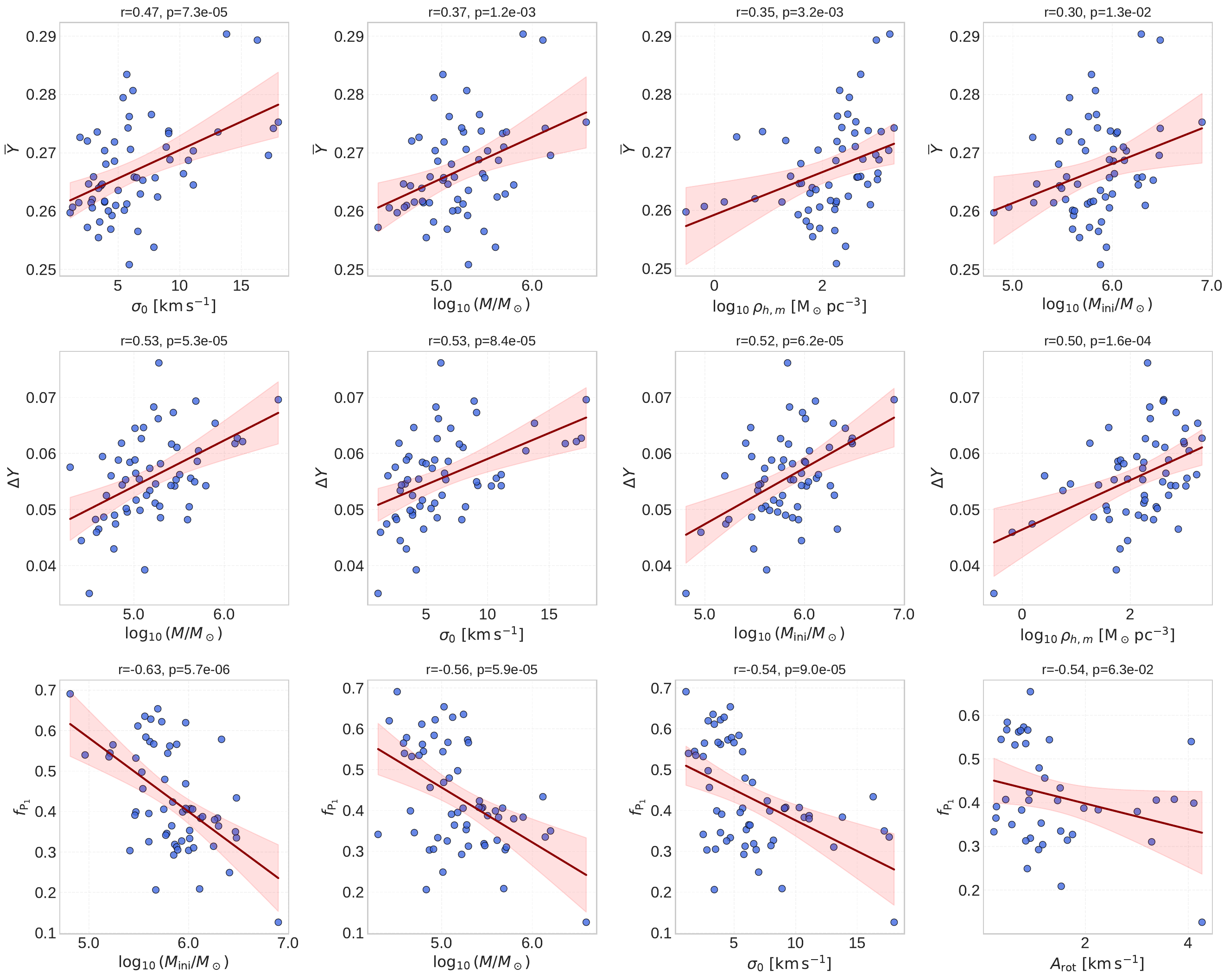}
    \caption{The four strongest correlations for each of the three MP parameters derived in this work: mean helium abundance $  \overline{Y}  $ (top row), helium spread $  \Delta Y  $ (middle row), and first-population fraction $  f_{\rm P1}  $ (bottom row). Each panel shows the Pearson correlation coefficient $  r  $ and two-sided $  p  $-value. The red line indicates the ordinary least-squares linear fit, with the shaded red region representing the 95\% confidence interval on the fit. Structural and dynamical parameters are taken from \cite{Baumgardt18} and include central velocity dispersion $  \sigma_0  $, present-day mass $  M  $, central luminosity density $  \rho_{h,m}  $, initial mass $  M_{\rm ini}  $ (derived following \cite{Baumgardt03}), and rotational velocity $  A_{\rm rot}  $.}
    \label{fig:top5}
\end{figure}

\subsection{Multiple Populations and Correlations with Cluster Parameters}
\label{sec:correlations}

One of the most robust ways to validate the physical reliability of our photometrically derived MP properties is to test whether they reproduce the primary correlations established in the literature between MP-related quantities and cluster structural and dynamical parameters. In particular, previous studies have consistently demonstrated that helium enrichment both the mean abundance $\overline {Y}$ and the internal spread $\Delta Y$ increases with cluster mass and central concentration, while the first-population fraction $ f_{\rm P1} $ decreases accordingly. Recovering these well-documented trends with our framework provides strong, independent confirmation of the validity of the derived helium abundances, helium spreads, and population fractions.

To perform this test, we adopt the comprehensive dataset of present-day masses, structural parameters, central velocity dispersions ($  \sigma_0  $), and central densities ($  \rho_{h,m}  $) from \cite{Baumgardt18} and associated references, supplemented by initial masses ($  \log M_{\rm ini}  $) derived from cluster orbits, present-day masses, and the prescriptions in Eqs.~(10) and (12) of \cite{Baumgardt03}. The key trends are illustrated in Figure~\ref{fig:top5}, which presents the four strongest correlations for each of the three MP quantities ($  \overline{Y}  $, $  \Delta Y  $, and $  f_{\rm P1}  $).

In agreement with prior photometric and spectroscopic investigations \citep{Milone17,Lagioia18,Milone18,ReviewBastian18}, both $  \Delta Y  $ and $  \overline{Y}  $ exhibit clear positive correlations with present-day cluster mass, initial mass ($  \log M_{\rm ini}  $), central velocity dispersion $  \sigma_0  $, and central density $  \rho_{h,m}  $. Although the slopes are relatively shallow (e.g., $  \sim 0.001  $ dex per km s$  ^{-1}  $ for $  \sigma_0  $, and $  \sim 0.007  $--$0.010$ dex per dex in mass or initial mass), these relations are among the most statistically significant in the dataset. In contrast, $  f_{\rm P1}  $ shows strong negative correlations with the same parameters, most notably with initial mass ($  r_{\rm weighted} \approx -0.63  $). While our absolute $  f_{\rm P1}  $ values are systematically offset from some literature estimates owing to methodological differences (see Section~\ref{sec:MP_fractions}), the relative trends remain robust and fully consistent with previous findings.

The successful recovery of these well-established, mass-dependent trends in close agreement with independent studies provides strong validation of our photometric two-population isochrone-fitting framework and confirms the reliability of the derived MP properties.

\end{document}